\documentclass[11pt]{article}
\usepackage{epsf,latexsym,amsmath,amssymb}

\begin{document}

%
%
%
%


\newcommand{\nc}{\newcommand}



\newcommand{\bra}[1]{\langle\; #1\; |}
\newcommand{\ket}[1]{|\; #1\; \rangle}
\newcommand{\Bra}[1]{\Big\langle\; #1\; \Big\lvert}
\newcommand{\Ket}[1]{\Big\rvert\; #1\; \Big\rangle}
\newcommand{\brap}[1]{{\langle\; #1\; |}^{\prime}}
\newcommand{\ketp}[1]{{|\; #1\; \rangle}^{\prime}}


\nc{\isk}[1]{|\; #1\; \rangle\rangle}
\nc{\isb}[1]{\langle\langle\; #1\; |}
\nc{\Isk}[1]{\Big\rvert\; #1\; \Big\rangle\Big\rangle}
\nc{\Isb}[1]{\Big\langle\Big\langle\; #1\; \Big\lvert}


\newcommand{\be}{\begin{equation}}
\newcommand{\ee}{\end{equation}}
\newcommand{\bes}{\begin{equation*}}
\newcommand{\ees}{\end{equation*}}
\newcommand{\ben}{\begin{eqnarray}\displaystyle}
\newcommand{\een}{\end{eqnarray}}
\newcommand{\bea}{\begin{eqnarray*}\displaystyle}
\newcommand{\eea}{\end{eqnarray*}}
\newcommand{\refb}[1]{(\ref{#1})}


\catcode`\@=11
%
%
%
\def\@citex[#1]#2{%
\if@filesw \immediate \write \@auxout {\string \citation {#2}}\fi
\@tempcntb\m@ne \let\@h@ld\relax \def\@citea{}%
\@cite{%
  \@for \@citeb:=#2\do {%
    \@ifundefined {b@\@citeb}%
      {\@h@ld\@citea\@tempcntb\m@ne{\bf ?}%
      \@warning {Citation `\@citeb ' on page \thepage \space undefined}}%
      {\@tempcnta\@tempcntb \advance\@tempcnta\@ne%
      \@tempcntb\number\csname b@\@citeb \endcsname \relax%
      \ifnum\@tempcnta=\@tempcntb 
        \ifx\@h@ld\relax%
          \edef \@h@ld{\@citea\csname b@\@citeb\endcsname}%
        \else%
          \edef\@h@ld{\ifmmode{-}\else--\fi\csname b@\@citeb\endcsname}%
        \fi%
      \else
        \@h@ld\@citea\csname b@\@citeb \endcsname%
        \let\@h@ld\relax%
      \fi}%
    \def\@citea{,\penalty\@highpenalty\,}%
  }\@h@ld
}{#1}}

%
\def\@citeb#1#2{{[#1]\if@tempswa , #2\fi}}
%
%
\def\@citeu#1#2{{$^{#1}$\if@tempswa , #2\fi }}
%
%
\def\@citep#1#2{{#1\if@tempswa , #2\fi}}

%
%
\def\bcites{         
        \catcode`\@=11
        \let\@cite=\@citeb
        \catcode`\@=12
}

\def\upcites{         
        \catcode`\@=11
        \let\@cite=\@citeu
        \catcode`\@=12
}

\def\plaincites{      
        \catcode`\@=11
        \let\@cite=\@citep
        \catcode`\@=12
}


\newcommand{\fn}{\footnote}
\newcommand{\sectiono}[1]{\section{#1}\setcounter{equation}{0}}


\newcommand{\bec}{\begin{center}}
\newcommand{\eec}{\end{center}}
\newcommand{\cl}{\centerline}


\catcode`\@=11
\renewcommand{\theequation}{\thesection.\arabic{equation}}
\@addtoreset{equation}{section}
\catcode`\@=12


\def\appendix{{\section*{Appendix}}\let\appendix\section%
        {\setcounter{section}{0}
        \gdef\thesection{\Alph{section}}}\section}

\nc{\beb}{\bigl[\begin{smallmatrix}}
\nc{\eeb}{\end{smallmatrix}\bigr]}
\nc{\bep}{\begin{pmatrix}}
\nc{\eep}{\begin{pmatrix}}


\newcommand{\non}{\nonumber}
\newcommand{\bh}{\bar{h}}
\nc{\etal}{\eta_L}
\nc{\etar}{\eta_R}
\nc{\imply}{\Rightarrow}
\nc{\chat}{\hat{c}}
\nc{\N}{{\sf N}}
\nc{\NN}{{\cal N}}
\nc{\SSS}{{\cal S}}
\nc{\NS}{\NN\SSS}
\nc{\RA}{{\cal R}}
\nc{\OO}{{\cal O}}
\nc{\PP}{{\cal P}}
\nc{\SP}{{\sf P}}
\nc{\UU}{{\cal U}}
\nc{\CY}{\text{CY}}
\nc{\ox}{\otimes}
\nc{\x}{\times}
\nc{\w}{\wedge}
\nc{\W}{\bigwedge}
\nc{\p}{\partial}
\nc{\pbar}{\bar{\partial}}
\nc{\MM}{{\sf M}}
\nc{\wt}{\widetilde}
\nc{\wh}{\widehat}
\nc{\vp}{\varphi}
\nc{\ep}{\epsilon}
\nc{\vep}{\varepsilon}
\nc{\vtheta}{\vartheta}
\nc{\RR}{\mathbf{R}}
\nc{\RE}{{\cal R}}
\nc{\Z}{\mathbf{Z}}
\nc{\ZZ}{\mathbf{Z_2}}
\nc{\RP}{\mathbb{R}\mathbb{P}}
\nc{\viz}{{\em{viz. }}}
\nc{\eg}{{\em{e.g. }}}
\nc{\ie}{{\em{i.e. }}}
\nc{\ol}{\overline}
\nc{\cplx}{\mathbf{C}} 
\nc{\zbar}{\bar{z}}
\nc{\wbar}{\ol{w}}
\nc{\noi}{\noindent}
\nc{\eights}{\boldsymbol{8}_{\boldsymbol{\rm s}}}
\nc{\eightc}{\boldsymbol{8}_{\boldsymbol{\rm c}}}
\nc{\sla}{(S_L)_{\alpha}}
\nc{\slb}{(\ol{S}_L)^{\dot{\alpha}}}
\nc{\sra}{(S_R)_{\beta}}
\nc{\srb}{(\ol{S}_R)^{\dot{\beta}}}
\nc{\wil}{(-1)^{F_L}}
\nc{\wi}{(-1)^{F}}
\nc{\II}{{\cal I}}
\nc{\g}{\mathfrak{g}}
\nc{\JJ}{{\cal J}}
\nc{\TO}{{\cal T}}
\nc{\KB}{{\cal K}}
\nc{\AN}{{\cal A}}
\nc{\MS}{{\cal M}}
\nc{\CC}{{\cal C}}
\nc{\HH}{{\cal H}}
\nc{\GG}{{\cal G}}
\nc{\one}{{\mathbf{1}}}
\nc{\tq}{\tilde{q}}
\nc{\tn}{\tilde{n}}
\nc{\tl}{\tilde{l}}
\nc{\tm}{\tilde{m}}
\nc{\ts}{\tilde{s}}
\nc{\Lbar}{\ol{L}}
\nc{\dps}{\displaystyle}
\nc{\Jbar}{\ol{J}}
\nc{\ibar}{\bar{i}}
\nc{\jbar}{\bar{j}}
\nc{\YY}{{\cal Y}}
\nc{\jhat}{\wh{j}}
\nc{\pjhat}{\wh{j^{\prime}}}
\nc{\nhat}{\wh{n}}
\nc{\pnhat}{\wh{n^{\prime}}}
\nc{\shat}{\wh{s}}
\nc{\jtilde}{\wt{j}}
\nc{\pjtilde}{\wt{j^{\prime}}} 
\nc{\ntilde}{\wt{n}}
\nc{\pntilde}{\wt{n^{\prime}}}
\nc{\stilde}{\wt{s}}
\nc{\pstilde}{\wt{s^{\prime}}}
\nc{\jprime}{j^{\prime}}
\nc{\nprime}{n^{\prime}}
\nc{\sprime}{s^{\prime}}
\nc{\sufusion}{\NN^{\,j}_{\;\;\;\jhat\,\pjhat}}
\nc{\ssufusion}{\NN^{\,\frac{k}{2}\,-\,j}_{\;\jhat\,\pjhat}}
\nc{\mmfusion}
{\NN^{\,j\,n\,s}_{\;\jhat\,\nhat\,\shat\;;\;\pjhat\,\pnhat\,\pshat}}
\nc{\half}{\frac{1}{2}}
\nc{\threehalf}{\frac{3}{2}}
\nc{\tlh}{\tilde{h}}
\nc{\tbh}{\tilde{\bh}}

\nc{\bz}{\zbar}

\nc{\ntil}{\wt{n}_r}
\nc{\nhatr}{\nhat_r}
\nc{\nprimer}{\nprime_r}
\nc{\ndiff}{\frac{\ntil\,-\,\nhatr}{2}}
\nc{\win}{(-1)^{\,\nhatr}}
\nc{\twin}{(-1)^{\,\ntil}}

\nc{\jhatr}{\jhat_r}
\nc{\jprimer}{\jprime_r}
\nc{\jtilder}{\jtilde_r}


\nc{\wij}{(-1)^{\,j_r}}

\nc{\stil}{\wt{s}_r}
\nc{\shatr}{\shat_r}
\nc{\sprimer}{\sprime_r}
\nc{\sdiff}{\frac{\stil\,-\,\shatr}{2}}
\nc{\wis}{(-1)^{\,\shatr}}
\nc{\twis}{(-1)^{\,\stil}}

\nc{\ndelone}{\delta^{\,6}_{\,\nprimer\,,\;\ndiff\,-\,\nu_0}}
\nc{\ndeltwo}{\delta^{\,6}_{\,\nprimer\,,\;\ndiff\,+\,3\,-\,\nu_0}}
\nc{\ndelthree}{\delta^{\,2}_{\,\ntil\,+\,\nhatr}}
\nc{\ndelfour}{\delta^{\,2}_{\,\ntil\,+\,\nhatr\,+\,1}}
\nc{\ndelfive}
{\delta^{\,6}_{\,\nprimer\,,\;\frac{\ntil\,-\,\nhatr\,+\,3}{2}\,-\,\nu_0}}
\nc{\ndelsix}
{\delta^{\,6}_{\,\nprimer\,,\;\frac{\ntil\,-\,\nhatr\,+\,9}{2}\,-\,\nu_0}}
\nc{\ndelseven}
{\delta^{\,6}_{\,\nprimer\,,\;\frac{\ntil\,-\,\nhatr\,-\,3}{2}\,-\,\nu_0}}
\nc{\ndeleight}{\delta^{\,2}_{\,\nprimer\,+\,\ntil}}
\nc{\ndelnine}
{\delta^{\,6}_{\frac{\nprimer}{2}\,,\;\frac{\ntil}{2}\,-\,\nhatr\,-\nu_0}}
\nc{\ndelten}
{\delta^{\,6}_{\frac{\nprimer}{2}\,,\;\frac{\ntil}{2}\,-\,\nhatr\,
+\,3\,-\nu_0}}
\nc{\ndeleleven}{\delta^{\,6}_{\,\nprimer\,+\,\ntil\,+\,1}}
\nc{\ndeltwelve}
{\delta^{\,6}_{\frac{\nprimer}{2}\,,\;\frac{\ntil}{2}\,-\,\nhatr\,
+\,\threehalf\,-\nu_0}}
\nc{\ndelthirteen}
{\delta^{\,6}_{\frac{\nprimer}{2}\,,\;\frac{\ntil}{2}\,-\,\nhatr\,
+\,\frac{9}{2}\,-\nu_0}}
\nc{\ndelfourteen}
{\delta^{\,6}_{\frac{\nprimer}{2}\,,\;\frac{\ntil}{2}\,-\,\nhatr\,
-\,\threehalf\,-\nu_0}}

\nc{\sdelone}{\delta^{\,4}_{\,\sprimer\,,\;\sdiff\,-\,\nu_0\,-\,2\,\nu_r}}
\nc{\sdeltwo}
{\delta^{\,4}_{\,\sprimer\,,\;\sdiff\,+\,2\,-\,\nu_0\,-\,2\,\nu_r}}
\nc{\sdelthree}{\delta^{\,2}_{\,\stil\,+\,\shatr}}
\nc{\sdelfour}
{\delta^{\,4}_{\,\sprimer\,,\;\frac{\stil\,-\,\shatr\,+\,2}{2}\,
-\,\nu_0\,-\,2\,\nu_r}}
\nc{\sdelfive}
{\delta^{\,4}_{\,\sprimer\,,\;\frac{\stil\,-\,\shatr\,-\,2}{2}\,
-\,\nu_0\,-\,2\,\nu_r}}
\nc{\sdelsix}{\delta^{\,2}_{\sprimer\,+\,\stil}}
\nc{\sdelseven}
{\delta^{\,4}_{\,\frac{\sprimer}{2}\,,\;\frac{\stil}{2}\,-\,\shatr\,-\,
\nu_0\,-\,2\,\nu_r}}
\nc{\sdeleight}
{\delta^{\,4}_{\,\frac{\sprimer}{2}\,,\;\frac{\stil}{2}\,-\,\shatr\,+\,2
\,-\,\nu_0\,-\,2\,\nu_r}}
\nc{\sdelnine}
{\delta^{\,4}_{\,\frac{\sprimer}{2}\,,\;\frac{\stil}{2}\,-\,\shatr\,+\,1
\,-\,\nu_0\,-\,2\,\nu_r}}
\nc{\sdelten} 
{\delta^{\,4}_{\,\frac{\sprimer}{2}\,,\;\frac{\stil}{2}\,-\,\shatr\,-\,1
\,-\,\nu_0\,-\,2\,\nu_r}}
\nc{\sdeleleven} 
{\delta^{\,4}_{\,\frac{\sprimer}{2}\,,\;\frac{\stil}{2}\,-\,\shatr\,-\,3
\,-\,\nu_0\,-\,2\,\nu_r}}

\nc{\alphat}{\wt{\alpha}}
\nc{\alphah}{\wh{\alpha}}

\nc{\normh}{\frac{1}{\kappa^{'A}_{\wh{\alpha}}}}
\nc{\normt}{\frac{1}{\kappa^{'A}_{\wt{\alpha}}}}

\nc{\norm}{\frac{1}{\kappa^{'A}_{\alphat}\,\kappa^A_{'\alphah}}}

\nc{\no}{n_1^r}
\nc{\nt}{n_2^r}
\nc{\so}{s_1^r}
\nc{\st}{s_2^r}



\nc{\va}{(\tau\,,\,\frac{z}{5}\,,\,0)}
\nc{\vah}{\Big(\frac{\tau}{2}\,,\,\frac{z}{5}\,,\,0\Big)}



\nc{\xto}{\xrightarrow}


%
%
%
%

\def\Bbb{\mathbb}
\def\BZ{\mbox{$\Bbb Z$}} \def\BR{\mbox{$\Bbb R$}}
\def\BC{\mbox{$\Bbb C$}} \def\BP{\mbox{$\Bbb P$}}
\def\CP{\BC\BP} 


\setcounter{secnumdepth}{3}
\setcounter{page}{1}

\begin{titlepage}
\begin{flushright}
{\tt hep-th/0306257}\\
{\tt RUNHETC-2003-19}
\end{flushright}
\begin{center}
{\Large\bf Crosscaps in Gepner Models and Type IIA Orientifolds}
\end{center}
\vspace*{.7cm}
\begin{center}
{\large\rm Suresh Govindarajan} \\
{\it Department of Physics\\
Indian Institute of Technology, Madras \\
Chennai 600 036 INDIA} \\
Email: {\tt suresh@chaos.iitm.ernet.in} \\
and \\[6pt]
{\large\rm Jaydeep Majumder} \\
{\it Department of Physics and Astronomy,\\ Rutgers University,
 Piscataway, NJ 08855-0849 USA}    \\
Email: {\tt jaydeep@physics.rutgers.edu} 
\end{center}
\begin{abstract}
As a first step to a detailed study of orientifolds of Gepner models
associated with Calabi-Yau manifolds, we construct crosscap states
associated with anti-holomorphic involutions (with fixed points)
of Calabi-Yau manifolds. We argue that these orientifolds
are dual to M-theory compactifications
on (singular) seven-manifolds with $G_2$ holonomy. Using the spacetime picture
as well as the M-theory dual, we discuss aspects of the orientifold that
should be obtained in the Gepner model. This is illustrated for the
case of the quintic.

\end{abstract}
\end{titlepage}
\newpage
\tableofcontents

\sectiono{Introduction and Summary}\label{bs1}

M-theory compactifications involving {\em compact} seven-dimensional 
manifolds of $G_2$
holonomy are of interest because they lead to compactifications with
minimal
i.e., ${\cal N}=1$ supersymmetry in four 
dimensions\cite{9812205,0011089,0101206,0103011,0107177,0108165,0109152,0203256,0302021}(See 
ref.\cite{BOBBY} for a review and a list of references). 
Other options to
obtain minimal supersymmetry include compactifications of the heterotic
string on CY threefolds and F-theory on CY fourfolds. In some cases,
all these might be related to each other by some form of duality.

It is also useful to consider M-theory to be the strong coupling limit
of the type IIA string theory. A large class of $G_2$ manifolds,
the {\em Joyce manifolds}, have
been constructed by Joyce using orbifold methods\cite{JOYCE1,JOYCE2}. 
One particular class, called
``barely $G_2$ manifold'', is obtained by orbifolding the manifold 
$\CY_3\ox S^1$. We shall consider the $\CY_3$ to be at the Gepner point of
its moduli space. By generalising an
argument due to Kachru and McGreevy\cite{KM}, we will argue that 
barely $G_2$
Joyce manifolds appear as the strong coupling limit of certain
 $\CY_3$ orientifolds at the Gepner point. We will study
these orientifolds using several different approaches. These fall
into two classes: the spacetime approach and the worldsheet 
approach. Within the latter, we take the boundary state 
approach. Construction of orientifolds \viz 
crosscaps states and computing Klein bottle amplitudes out of those crosscaps 
in Gepner model will be the main theme of this paper. For models of 
phenomenological interest, \eg, type IIA orientifolds on $T^6/(\ZZ\ox\ZZ)$ 
and  their
lift to M-theory on $G_2$ manifolds, see refs.\cite{CSU1,CSU2,CSU3}. 

Orientifolds\cite{SAGNOTTI,CALLAN,POLCHINSKICAI,HORAVA} of compact 
Calabi-Yau manifolds has been a subject of recent 
interest\cite{Acharya:2002ag,0206038,0304209,0305021}.\fn{For type II 
theories on toroidal orientifolds in pre-D-brane era, see refs.\cite{BS}.}
Orientifolds of WZW 
CFT 
and coset CFTs had been investigated in 
refs.\cite{BRUNNER,HIKIDA,0110267,HALPERN}.\fn{See the 
very recent paper ref.\cite{FUCHS} which relate TFT and RCFT on 
unoriented worldsheet.}
The most thorough 
investigation of orientifolds of generic rational conformal field 
theory, \eg, bosonic parafermions, $\NN=2$ minimal 
models and Calabi-Yau from the CFT point of view had been done in 
refs.\cite{BH1,BH2}. For earlier attempts on orientifolds of Gepner 
models, mainly type I theory on Gepner models, see 
refs.\cite{9806131,9607229}. D-branes in coset CFTs had been studied in 
ref.\cite{MMS}; Study of D-branes in Gepner models was initiated in 
\cite{Recknagel:1997sb} and 
carried out thoroughly for the case of quintic in ref.\cite{9906200}.

Our method will be example oriented because we believe the general procedure
for constructing orientifolds in rational conformal field theory(RCFT) can be 
best understood if we focus on particular examples. We found that Gepner 
models constructed out of level $k_r$ minimal model with $k_r=\text{even}$ for
some $r$, is quite intricate as compared to the $k_r=\text{odd}$ counterpart.
This is also reflected in the geometric limit as the anti-holomorphic
involutions and their fixed point sets are
much richer. For simplicity, in this 
paper
we mainly deal with Gepner models with $k_r=\text{odd}$ for all $r$, 
reserving the detailed investigation on $k_r=\text{even}$ models for 
future work. 
Moreover, within $k_r=\text{odd}$ Gepner models we pick those with only one
K\"ahler modulus; the canonical choice for such $\CY_3$ is the quintic. 

We start by considering M-theory on a barely $G_2$ manifold 
$\dfrac{\CY_3\ox S^1}{\sigma_0\cdot I_1}$ where $\sigma_0$ is an 
antiholomorphic
involution\fn{The image of $\sigma_0$ in type IIA theory will be denoted as 
$\sigma$.} of $\CY_3$ and $I_1$ is the inversion of the circle $S^1$. For
$\CY_3$ we consider the Fermat's quintic hypersurface and its corresponding
Gepner model $(k=3)^5$. As we go to the $g_s\to 0$ limit, 
the circle $S^1$ shrinks
in size and we are left with type IIA theory on the 
orientifold\cite{KM,SEN,JM}
$\CY_3/\Omega\cdot\sigma\wil$, where $\Omega$ is the worldsheet parity and
$F_L$ denote left-moving spacetime fermion number\fn{When we consider the 
Gepner model of such $\CY_3$, the orientifold group will also contain
the discrete symmetry group of the Gepner model.} . Suppose, somebody asks at
this point the following question : what are the crosscap states in this 
theory? Answering this question was our main motivation and subject of this
paper.

We now outline the organization of our paper. We start in section 
\ref{bs4} with the geometric 
approach first by working out the massless spectra of M-theory on barely
$G_2$ manifold. In section \ref{scas1}, we check this spectra by working
in the worldsheet approach. Here we work out the action of the 
orientifold projection operator on each massless state
in $\NN=2$, $c=9$ superconformal algebra and work out the massless fields 
in the projected theory. It agrees with the analysis of section \ref{bs4}
and also with the rules in ref.\cite{Vafa:1995gm}, 
given years ago by working on 
the geometric side.

Section \ref{bs2} contain the 
basic set up for treating unoriented strings in RCFT with and without simple 
currents. It mainly
summarizes the foundations laid in refs.\cite{CARDY,PSS}. The reader is 
requested to go through this chapter first, since it contains all the major
formulae and forms the backbone for subsequent chapters. Other informations
relevant to the section \ref{bs2} can be found in appendices \ref{A1}, 
\ref{A2} and \ref{A3}.

Sections \ref{ns1} and \ref{qs1} form the heart of the paper. As a warm-up
and to prepare the groundwork for the orientifold of the quintic model, 
we discuss the type IIA 
orientifold on $(k=1)^3$ Gepner model in section \ref{ns1}. The orientifold
group can be enlarged by incorporating the discrete symmetries of the Gepner
model. This can be found in section \ref{nsss1.2}. The most important 
formula
for the $P$-matrix is given in appendix \ref{A4s3}. The explicit 
expressions
for the crosscaps in {\em internal} CFT are given in section \ref{nsss1.3}.
To get a physical insight for the Klein bottle amplitude of this model, we
first discuss the spectral-flow invariant orbits for $(k=1)^3$ Gepner model.
In section \ref{nss1.3}, we compute the Klein bottle amplitude for this model
by using the crosscaps of section \ref{nsss1.3} and express the answer in 
terms of those spectral-flow invariant orbits. The {\em master formula} for 
the Klein bottle amplitude can be found in eqn.\refb{nKB} in section 
\ref{nsss1.3}. Though these formulae for crosscaps and Klein bottle amplitudes
are derived for $(k=1)^3$ Gepner model, it is
general enough in that it can be applied to any Gepner model with all levels
$k_r$ being {\em odd}; in particular, it can be applied to the quintic in a 
straightforward way. As expected, we get the Klein bottle amplitude for this
model to be proportional to the {\em massless} orbit of the model. It was
quite satisfying and show that the abstract RCFT construction is on the right
track. This warm-up example gave us enough confidence to apply this technology
directly to the quintic Gepner model. The explicit formulae for the 
characters of $k=1$ minimal model can be found in appendix \ref{A6}.

In section \ref{qs1}, we discuss our main goal, \ie the crosscaps and Klein
bottle amplitudes in quintic. It is easy now to write down the crosscaps using
the method of section \ref{nsss1.3} and is given in eqn.\refb{q5c}. Before
computing the Klein bottle amplitude in $(k=3)^5$ model, we write down the 
spectral flow invariant orbits of this model. Though we searched hard in the 
literature, we think this is for the first time the spectral flow invariant
orbits are being worked out. There are many of them but we mention the most
important of them in section \ref{qss1.3}.  As a cross check, we compare it 
with the characters of $c=9$ algebra --- the worldsheet algebra relevant for
manifolds of $SU(3)$ holonomy\cite{EOTY,ODAKE}. After discussing the 
orientifold
group of this model, we compute the Klein bottle amplitude in section 
\ref{qss1.4}. The answer is given in eqn.\refb{q6}; as expected it only 
involves the massless graviton and self-conjugate matter orbits. For further
details, see the relevant section.  We gather the formulae for the characters
and {\em string functions} for $k=3$ minimal model in the appendix \ref{A7}.

We conclude in section \ref{con1}. In appendix \ref{A4} we gather various 
important formulae for $\NN=2$ minimal model. 

As we were finishing this project, we received a preprint\cite{SH}, which
has overlaps with our section \ref{cs1}. They 
proposed a crosscap similar to our eqn.\refb{gepnercrosscap} in section 
\ref{cs1}, though we proposed it much earlier\cite{sureshpascos} 
and this issue had 
been clearly addressed in our PASCOS '03 conference report 
ref.\cite{Pascos}.

\sectiono{Geometric analysis}\label{bs4}

We shall briefly discuss the Joyce's construction of seven-dimensional 
manifolds with barely $G_2$-holonomy. Consider the seven-dimensional orbifold 
$X$
given by the $\BZ_2$ action on the manifold $M\times S^1$, where $M$ is
a Calabi-Yau manifold admitting an anti-holomorphic involution ${\sigma}$
and the $\BZ_2$ is given by an inversion $I_1$
of the $S^1$ and the anti-holomorphic
involution ${\sigma}$. Thus,
\begin{equation}\label{b2.47}
X = \left(M\times S^1\right)/{\sigma}\cdot I_1\quad.
\end{equation}
The anti-holomorphic involution $\sigma$ is an isometry of the Calabi-Yau
manifold; it acts on its K\"ahler form $J$ and the holomorphic 3-form 
$\Omega^{(3)}$
as :
\begin{equation}\label{b2.47a}
\sigma(J)\,=\,-\,J\,,\qquad 
\sigma(\Omega^{(3)})\,=\,e^{i\,\theta}\,\Omega^{(3)}\,,
\end{equation}
where $\theta$ is a real phase. This action of $\sigma$ on $X$ and the 
inversion $I_1$ on the circle $S^1$ preserves the 3-form
\begin{equation}\label{b2.47b}
\phi\,=\,J\wedge d\,x\,+\,\Re{(e^{i\theta/2}\,\Omega^{(3)})}
\end{equation}
on $M$ and equips it with a $G_2$ structure; hence the projection 
$\sigma\cdot I_1$ preserves $\NN\,=\,1$ supersymmetry in $D\,=\,4$.

One distinguishes two cases: (i) the involution ${\sigma}$ has no fixed
points and (ii) the involution ${\sigma}$ has a submanifold $\Sigma\in M$
as its fixed point. For case (i), $X$ is a non-singular manifold with
$G_2$ holonomy while for case (ii) $X$ is a singular manifold with
$G_2$ holonomy. Joyce shows that if $b_1(\Sigma)>0$, the singularity
can be blown up to obtain a non-singular manifold with $G_2$-holonomy.
When $b_1(\Sigma)=0$, the manifold has a non-smoothable singularity.

For purposes of compactifications of M-theory on manifolds with
$G_2$ holonomy, singularities are a virtue -- they provide us with
examples of  ${\cal N}=1$ supersymmetric
compactifications in four dimensions with non-abelian gauge symmetry
as well as chiral fermions, both of phenomenological 
interest\cite{0107177,0109152}.

Anti-holomorphic involutions also make an appearance in a different context.
They provide us with a large class of special Lagrangian (sL) submanifolds
of Calabi-Yau threefolds. Thus, $\Sigma$ can be chosen to be  sL
submanifold of the CY threefold. In such cases, we will argue, extending
an argument due to Kachru and McGreevy\cite{KM}, that the type IIA dual of
the M-theory compactification on $X$ is a type IIA orientifold of $M$,
with the $S^1$ playing the role of the eleven-dimensional circle:
$$
\boxed{
{\rm M-theory\ on\ X} \stackrel{\rm dual}{\longleftrightarrow} {\rm
Orientifold\ of\ type\ IIA\ on\ M}
}
$$
The appearance of non-abelian gauge symmetry is easily seen in the
orientifold due to the addition of D6-branes wrapping $\Sigma$ and
extending in the non-compact spacetime in order to cancel the RR-tadpoles
due to the presence of orientifold 6-planes. Suppose $\Sigma=L_1 + L_2$,
where $L_1$ and $L_2$ are two sL submanifolds which preserve the same
supersymmetry and intersect each other at a point. 
Then, following ref. \cite{Berkooz:1996km}, one
expects the appearance of chiral fermions at the points of intersection.

We need to map the inversion of the eleven dimensional circle
to a suitable orientifold action in order to obtain the correct
type IIA dual for the M-theory compactification. Let us first consider
the type IIA string in flat spacetime. Inversion of an odd
number of coordinates is not a symmetry -- it can however be made a
symmetry by including the simultaneous action of worldsheet parity
and possibly $(-)^{F_L}$. The precise choice depends on the number of
coordinate inversions. From the analysis of Sen\cite{SEN,SENREV}, one 
finds that
the correct choices are given by 
\begin{equation}\label{b2.48}
\BR^{9-p}/{\cal I}_{9-p}\cdot\Omega\cdot g\quad,
\end{equation}
where ${\cal I}_{9-p}$ reverses the sign of all coordinates on
$\BR^{9-p}$, $\Omega$ is the worldsheet parity operation and
\begin{equation}\label{b2.48a}
g=\left\{ \begin{matrix}
1 & {\rm for}\ (9-p)=0,1  \mod 4 \cr
                  (-)^{F_L}& {\rm for}\ (9-p)= 2,3 \mod 4 
\end{matrix}\right.\quad.
\end{equation}
The origin is the fixed point of the orientifold action and is the
location of an orientifold $p$-plane. The RR-charge of the orientifold
plane can be (locally) cancelled by the addition of $32/2^{9-p}$ Dp-branes
in the convention that one adds 32 D9-branes to obtain type I theory 
as a type IIB orientifold. The enhanced gauge symmetry is SO$(32/2^{9-p})$
for an SO-type orientifold plane ($O^+$ in Witten's convention).

\subsection{The quintic hypersurface in $\BP^4$}\label{bss4.1}

The most studied example of a compact CY threefold is given by
the quintic hypersurface in $\BP^4$. The Fermat quintic $M_{\rm FQ}$
is given by the hypersurface given by equation
\begin{equation}\label{b2.49}
z_1^5 + z_2^5 + z_3^5 + z_4^5 + z_5^5 =0\quad,
\end{equation}
where $z_i$ are homogeneous coordinates of $\BP^4$. The fixed point of
the anti-holomorphic involution 
\begin{equation}\label{b2.49a}
{\sigma}:\qquad z_i\rightarrow \bar{z}_i\quad,\quad i=1,\ldots,5\quad,
\end{equation}
is an $\BR\BP^3$, which is a sL submanifold of the Fermat 
quintic\cite{Becker:1995kb}.
This submanifold is also the base of
SYZ $T^3$-fibration of the Fermat quintic\cite{SYZ}. The anti-holomorphic
involution corresponds to the inversion of all three circles of the
$T^3$-fibre. From equation (\refb{b2.48}), the
anti-holomorphic involution can be made a symmetry of type IIA string
by including worldsheet parity with $(-)^{F_L}$. Thus, the duality 
with M-theory can be made precise for this example
\begin{equation}\label{b2.50}
{\rm M-theory\ on\ }M_{\rm FQ}\times S^1/\sigma \cdot I_1
 \stackrel{\rm dual}{\longleftrightarrow} {\rm
Type\ IIA\ on\ M_{\rm FQ}/{\sigma}\cdot\Omega\cdot(-)^{F_L}}
\end{equation}
Tadpole considerations from the flat case suggest that
IIA orientifold will need the addition of four D6-branes wrapping the
$\BR\BP^3$ and filling spacetime. We will see that all this will be
consistent with our CFT analysis in the next section.

$\Sigma$ is actually one in a family of $5^4=625$ sL submanifolds of the
Fermat quintic, all of whom are $\BR\BP^3$'s. In fact the involution given in 
eqn.\refb{b2.49a} is a special case of eqn.\refb{b2.47a} with $\theta=0$. 
The most general anti-holomorphic involution with a non-vanishing $\theta$
are given by the following :
\begin{equation}\label{b2.51}
z_i \rightarrow \omega^{n_i} \bar{z}_i  \quad,
\end{equation}
where $\omega\,=\,e^{2\pi i/5}$ and  the identification
under a shift of all $n_i$ by one is understood. This is a trivial scaling of
all homogeneous coordinates. This freedom can be used
to set $n_5=0$. All these sL manifolds are further divided by
$\sum_i n_i\bmod 5$ -- this is related to the phase associated with
the sL condition. 

In ref.\cite{Partouche:2000uq,RRW}, all possible anti-holomorphic involution 
for the embedding
projective space $\CP^N$ had been classified. They are of two types. If $z^1$,
$z^2$,$\cdots$, $z^{N+1}$ denote the homogenous coordinates of $\CP^N$, then
these anti-holomorphic involutions are :
\ben\label{b2.51a}
A &\colon& (z^1,z^2,\cdots,z^{N+1})\,\xto{}\,(\bz^1,\bz^2,\cdots,\bz^{N+1})
\non\\
B &\colon& (z^1,z^2,\cdots,z^N,z^{N+1})\,\xto{}\,(-\bz^2,\bz^1,\cdots,
-\bz^{N+1},\bz^N)
\een
Here $B$ is defined only for {\it odd} $N$ and acts {\it freely} while $A$ has
a fixed set(for $z^i\,=\,\bz^i$). In our examples we mostly deal with $A$-type
involutions. Further, if the embedded Calabi-Yau surface is given as
$$
\sum_{i=1}^{N+1}\;z_i^{n_i}\,=\,0\,,
$$
then one can add another discrete symmetry 
\begin{equation}\label{b2.51b}
z^i\,\to\,\zeta_i\,z^i\,,\quad i\,=\,1\,,\cdots,\,N+1\,,
\end{equation}
to either of the involutions $A$ or $B$; here $\zeta_i$ is $n_i$-th root of 
unity. Here $\{\zeta_i\,;\,i\,=\,1\,,\cdots,\,N+1\}$ generate a subgroup 
of 
the 
automorphism group of the Calabi-Yau hypersurface or the associated 
Landau-Ginzburg superpotential or of the associated Gepner model. We shall 
give explicit examples of this dicrete symmetry groups in sections \ref{ns1} 
and \ref{qs1} in the context of $(k=1)^3$ and $(k=3)^5$ Gepner models. 
Together with the antiholomorphic involutions $A$ and worldsheet parity 
$\Omega$, we shall build the full orientifold group in these models.

\bigskip

\subsection{M-theory on $CY_3$ $\times S^1$: massless spectrum}\label{bss3.4}

The massless spectrum can be obtained from compactifying eleven-dimensional
supergravity on a $CY_3$ (as in Witten) and then further dimensionally
reducing on a circle. Alternately, one can compactify M-theory on
a circle to obtain the type IIA string and then compactify the IIA
string on the $CY_3$, $M$. We will pursue the second method. Of course, both
methods give the same spectrum. 

The compactification of M-theory on a circle gives rise to the
type IIA string whose bosonic spectrum is
\begin{equation}\label{b2.52}
\begin{matrix}
{\rm NSNS\ sector} & G_{\hat{\mu}\hat{\nu}}\quad 
B_{\hat{\mu}\hat{\nu}}\sim
C_{\hat{\mu}\hat{\nu}\theta}
\quad \Phi\sim G_{\theta\theta} \\
{\rm RR\ sector} & A^{(1)}_{\hat{\mu}}\sim G_{\hat{\mu}\theta}\quad 
A^{(3)}_{\hat{\mu}\hat{\nu}\hat{\rho}}\sim  C_{\hat{\mu}\hat{\nu}\hat{\rho}}
\end{matrix}
\end{equation}
where we have used hatted Greek indices to denote ten-dimensional
vectors and symbolically indicated the eleven-dimensional origin
of the various fields using $\theta$ to indicate the index for the
the $S^1$ and $C$ for the three-form gauge field.

On further compactification on the CY3 $M$
with $h^{1,1}(M)$ K\"ahler moduli and $h^{2,1}(M)$ complex moduli,
the NSNS fields decompose as
\begin{eqnarray}\label{b2.53}
G_{\hat{\mu}\hat{\nu}} &\longrightarrow& \left\{
\begin{matrix}
g_{\mu\nu} \\
g_{ij} = z^A\ \mu_{A\bar{i}\bar{j}} \\
g_{i\bar{j}} = v^k\ \omega_k 
\end{matrix}\right.\nonumber \\
B_{\hat{\mu}\hat{\nu}}&\longrightarrow& B_{\mu\nu}\ {\rm and}\ b^k\
\omega_{ki\bar{j}} \\
\Phi &\longrightarrow& \Phi \nonumber
\end{eqnarray}
where $\mu_{A}$ are the $h^{1,2}(M)$ Beltrami differentials 
that parametrise deformations of complex structure of $M$
and $\omega_k$ are a basis for $H^{1,1}(M)$.

In the RR sector, the one-form gauge field provides a gauge field 
(the graviphoton) $A_\mu$
in four dimensions.
The RR three-form $A^{(3)}$ can be decomposed as
\begin{equation}\label{b2.54}
A^{(3)}= \xi^A \alpha_A + \tilde{\xi}_B \beta^B + C^{(1)~k}
\omega_k\quad,
\end{equation}
$(\alpha_A,\beta^B)$ are a real basis for $H^3(M)$,
$A,B=1,\ldots,h^{2,1}(M)$ with $\alpha_A \wedge \beta^B = \delta_A^B$
and $\alpha_A\wedge\alpha_B=\beta^A\wedge\beta^B=0$. The holomorphic
three-form is taken to be $\alpha_0 + i \beta^0$.

The compactification of the type II string on a Calabi-Yau threefold 
gives rise to ${\cal N}=2$ supergravity in $d=4$ as its low-energy limit. 
Thus, one can organise the bosonic fields into multiplets of 
${\cal N}=2$ supergravity. The full massless spectrum is
\begin{enumerate}
\item {\bf The gravity multiplet}: the graviton, $g_{\mu\nu}$; 
the graviphoton, $A_\mu$ and two gravitini.
\item {\bf The universal hypermultiplet}: Four real scalars: the
dilaton $\Phi$, $b$ (after dualising $B_{\mu\nu}$), $\xi^0$,
$\tilde{\xi}_0$.
\item {\bf $h^{2,1}(M)$ Hypermultiplets}: The complex scalars $z^A$ and the 
real scalars $\xi^A$, $\tilde{\xi}_A$ and a Dirac fermion.
\item {\bf $h^{1,1}(M)$ Vector multiplets}: The gauge fields $C_{\mu}^{(1)~k}$,
complex scalars $t^k = b^k + i v^k$ and a Dirac fermion.
\end{enumerate}

\subsection{M-theory on Joyce manifolds: the massless spectrum}\label{bss3.5}

As we discussed earlier, Joyce manifolds are manifolds with $G_2$
holonomy that are obtained as an $\BZ_2$ orbifold of CY3 $\times S^1$.
The $\BZ_2$ is the combination of an anti-holomorphic involution of
the CY3 and inversion of the circle.

The orbifold projection breaks ${\cal N}=2$ down to ${\cal N}=1$. 
It is therefore
useful to decompose ${\cal N}=2$ multiplets into ${\cal N}=1$ multiplets. 
A hypermultiplet
breaks into two chiral multiplets, an ${\cal N}=2$ vector multiplet decomposes
into an ${\cal N}=1$ vector multiplet and a chiral multiplet. Finally, the 
${\cal N}=2$ gravity multiplet decomposes into the ${\cal N}=1$ supergravity 
multiplet and a ${\cal N}=1$ vector multiplet, 
which we will call the graviphoton mutiplet.

\noindent
The orbifold projection is as follows:
\begin{enumerate}
\item {\bf The gravity multiplet}: 
The graviphoton and its supersymmetric
partner which is one of the gravitini get projected out leaving 
behind a ${\cal N}=1$ supergravity multiplet.
\item {\bf The universal hypermultiplet}:
$b$ is projected out while the dilaton is projected in. One
linear combination of the $\xi^0$ and $\tilde{\xi}^0$ is
invariant under the anti-holomorphic involution. Thus, one is
left with a ${\cal N}=1$ chiral multiplet.
\item {\bf Hypermultiplets}: The inversion of the the $S^1$ does not affect
these fields and hence the anti-holomorphic involution alone plays
a role in the projection. This involution leaves invariant a `real'
set of fields: of the two real fields that make up $z^A$, one linear
combination is invariant and is projected in. The same story holds
for the $\xi^A$ and $\tilde{\xi}^A$.  They form $h^{2,1}$ chiral
multiplets.
\item {\bf Vector multiplets}: The inversion of the $S^1$ does
not affect 
the gauge fields $C_{\mu}^{(1)~k}$. The projection is 
determined wholly by whether the $(1,1)$ form $\omega_k$ is even or
odd under the anti-holomorphic involution. 
When $\omega_k$  are even, the vector
multiplet is projected in and otherwise they are projected out.
Among the scalars, $b^k$ is odd under the $S^1$ inversion
and thus it is projected in whenever the corresponding $(1,1)$ form 
$\omega_k$ is odd. Let $h_+^{1,1}$ be the number of even $(1,1)$ forms
and $h_-^{1,1}$ be the number of odd $(1,1)$ forms. Thus, one
has $h_+^{1,1}$ ${\cal N}=1$ vector multiplets and $h_-^{1,1}$ chiral multiplets
after the orbifold projection.
\end{enumerate}
Let us first consider this case when the oribifold has no fixed
points. The massless
spectrum is the same as the compactification
of M-theory on a smooth $G_2$ manifold $X$ with Betti 
numbers\cite{Papadopoulos:1995da,Harvey:1999as,Partouche:2000uq} 
$$
b_3(X) = h^{2,1}(M)+h_-^{1,1}(M)\quad,\quad
b_2(X) = h_+^{1,1}(M)\quad,
$$
i.e., there are $b_3(X)$ chiral multiplets and $b_2(X)$ abelian
vector multiplets in addition to the supergravity multiplet.

Our main focus will be on the cases when one has fixed points. There are
various possibilities if the seven-manifold $X$ is singular.
Non-abelian
(ADE) vector multiplets  with $A$--$D$--$E$-type gauge group $G$
arise in M-theory compactifications at singularities
of the form $\BC^2/\Gamma_{ADE}$. This is the simplest of all possibilities
and our case belongs to this category\fn{If $\Sigma$ is smooth and the
normal space to $\Sigma$ is a smoothly varying family of $A$--$D$--$E$ 
singularities, the (3 + 1)-dimensional low energy theory will be a theory
with gauge group $G$, without chiral matter. In this case the dimension of
the moduli space of the low energy theory is equal to $b_1(\Sigma)$. To get chiral matter, $\Sigma$ 
must be singular or it must pass through worse than orbifold singularities
of $X$. In that case the dimension of the moduli space of low energy
theory gets bigger than $b_1(\Sigma)$, since now one has to consider 
the moduli of complex
gauge connection along $\Sigma$; see ref.\cite{BOBBYMODULI} for more
details. We thank Bobby
Acharya for a discussion on this point.}. Thus, when the singularities are of
the form $\Sigma\times \BC^2/\Gamma_{ADE}$, we expect extra non-abelian
vector multiplets; here $\Sigma$ is a 3-manifold embedded in $X$. 
When $b_1(\Sigma)\neq0$, we expect to see the appearance
of $b_1(\Sigma)$ chiral multiplets which can be understood as the blowup
modes and according to Joyce, the singularity
can be smoothed out.
When $b_1(\Sigma)=0$, the singularity is non-smoothable and the
non-abelian vector multiplets remain.

We shall aim to reproduce these results in our analysis of the orientifold
which we have proposed as the dual to these M-theory compactifications
on these Joyce manifolds.

\sectiono{Orientifold projection in the CFT}\label{scas1}

In this section, we will work out the action of the orientifold group
on the vertex operators associated with various fields that appear in
the Calabi-Yau compactification of the type IIA string. 

\subsection{$N=2$ preliminaries}\label{scass1}
The worldsheet
has enhanced supersymmetry, $(2,2)$ with generators ($T_L(z)$,
$G^{\pm}_L(z)$, $J_L(z)$) and ($T_R(\bz)$, $G^{\pm}_R(\bz)$, $J_R(\bz)$) for
the left- and right-moving sector respectively, 
each generating an $N=2$ algebra. Primary states are thus labelled
by four numbers $(h_L,q_L,h_R,q_R)$, where $h_L$ ($h_R$) is the conformal
weight and $q_L$ ($q_R$)   is the $U(1)_L$ ($U(1)_R$) charge.
The $N=2$ superconformal algebra has  a parameter 
$a$($0\leq a<1$) related to boundary conditions on the fermionic generators
or equivalently on the moding of these generators\cite{warnerictp,greentasi}.
In the complex plane, 
the fermionic generators have integer moding $(a=0)$ in the Ramond sector and 
half-integer moding $(a=1/2)$ in the NS sector.

In the NS sector, there are a special class of primaries which satisfy
$h=|q|/2$. These are the {\em chiral} primaries with $h=+q/2$ and
{\em anti-chiral} with $h=-\frac{q}{2}$. In the Ramond sector, the ground states
in a unitary theory can be shown to have $h=c/24$. There is a mapping,
the {\em spectral flow},
which relates states in the Ramond sector to the NS sector.
In general, the $N=2$ algebras given by $a$ gets 
mapped to one with $(a+\eta)$ by means of spectral flow 
ith spectral parameter $\eta$. 
Under this action,
the primary given by $(h,q)$ gets mapped to the primary given by
\ben\label{SCA1}
h &\to& h_{\eta} = h \,-\, \eta\,q \,+\, {\frac{c}{6}}\,\eta^2 \non\\
q &\to& q_{\eta} = q_L \,-\, {\frac{c}{3}}\,\eta\ .
\een
From the above formulae, one can see that under a spectral flow
with $\eta=1/2$, chiral primaries get mapped to Ramond ground states
and under spectral flow with $\eta=-1/2$, anti-chiral primaries
get mapped to Ramond ground states.

Recall that there are two $N=2$ algebras, one each from the left- and right-
movers. We shall label the spectral parameters $\etal$ and $\etar$
respectively. The operators which generate spectral flow can be written
out explicitly on bosonisation. Let 
\begin{equation}\label{SCA16}
J_L(z) = i\,\sqrt{c/3}\,\p_z\,H_L(z)\quad{\rm and}\quad
J_R(\zbar) = i\,\sqrt{c/3}\,\p_{\zbar} \,H_R(\zbar)
\end{equation}
with the normalization given by $H_L(z)\,H_L(w) \sim -\,\ln{(z\,-\,w)}$
and $H_R(\zbar)\,H_R(\wbar) \sim -\,\ln{(\zbar\,-\,\wbar)}$. 
In terms of the bosons $H_L(z)$ and $H_R(\bz)$ the spectral flow operator 
corresponding to spectral parameters $(\etal,\etar)$ is
given by
\begin{equation}\label{SCAspectralflow}
U_{\,\etal\,,\etar}\,=\,e^{i\sqrt{\frac{c}{3}}(\etal\,H_L\,+\,\etar\,H_R)}
\end{equation}

We are interested in compactifications of type IIA on Calabi-Yau threefolds.
The Calabi-Yau sector has $c=9$ and the massless bosonic states arise from
two sectors: NSNS and RR. Spectral flow relates states in these sectors
and is closely related to supersymmetry in spacetime. We will now
tabulate the relevant states. The massless states in the NSNS sector
arise from the $(c,c)$, $(a,a)$ primaries [there are $h_{2,1}$ of these] 
and $(a,c)$ and $(c,a)$ primaries [there are $h_{1,1}$ of these]
with $h=1/2$. We used the obvious notation: $c$ for chiral primaries
and $a$ for anti-chiral primaries. Finally, there is the identity
operator which is both chiral and anti-chiral and has $h=0$.
In the following table, we set the notation for the operators
corresponding to these massless excitations(see for instance, 
refs. \cite{CALLAN,CECOTTI})
\begin{table}
\begin{center}
\begin{tabular}{|c|c|c|}\hline
Sector & NSNS $\xrightarrow{(\etal,\etar)}$ RR & Number \\ \hline
Identity & $1 \xrightarrow{(\frac12,-\frac12)~} \Sigma^0$ & $1$ \\ \hline
Identity & $1 \xrightarrow{(-\frac12,\frac12)~} \Sigma^{0\dagger}$ & $1$\\
\hline
Identity & $1 \xrightarrow{~(\frac12,\frac12)~} \Xi^0$ & $1$ \\ \hline
Identity & $1 \xrightarrow{(-\frac12,-\frac12)} \Xi^{0\dagger}$ & $1$\\
\hline
$(c,a)$  & $\Lambda^j \xrightarrow{(\frac12,-\frac12)~} \Sigma^{j}$ 
& $j=1,\ldots,h_{1,1}$\\ \hline
$(a,c)$&  $\Lambda^{j\dagger}\xrightarrow{(-\frac12,\frac12)~} 
 \Sigma^{j\dagger}$ & $j=1,\ldots,h_{1,1}$\\ \hline
$(c,c)$  & $\Pi^A \xrightarrow{~(\frac12,\frac12)~} \Xi^{A}$ 
& $A=1,\ldots,h_{2,1}$\\ \hline
$(a,a)$&  $\Pi^{A\dagger}\xrightarrow{(-\frac12,-\frac12)} 
 \Xi^{A\dagger}$ & $A=1,\ldots,h_{2,1}$\\ \hline
\end{tabular}
\end{center}
\caption{States and their relation under spectral flow}
\end{table}

\subsection{Vertex operators for various fields}\label{scass2}

The vertex operators for the various fields will be needed to study the
action of the orientifold group and implement in the projection. It is
natural to work in the $(-1,-1)$ picture for the NSNS vertex operators,
$(-\frac12,-\frac12)$ picture for RR vertex operators and the 
$(-\frac12,0)$ and $(0,-\frac12)$ pictures for the generators of 
spacetime supersymmetry. The vertex operators include pieces from the
spacetime sector as well as the ghost sector. The index $\mu$ is a vector
index of $SO(3,1)$ and $\alpha$($\dot{\alpha}$)
are indices for Weyl spinors with positive(negative)
chirality(see ref. \cite{WB} for notation). 
The (free) fields that make up the spacetime sector are 
$X^\mu$, $\psi_L^\mu$ and $\psi_R^\mu$.

\subsubsection{Supersymmetry Charges}\label{scasss1}

The type IIA compactification has $\NN=2$ supersymmetry in four dimensions.
The two supersymmetry charges arise from the R-NS and the NS-R sectors.
charges. Let us label them by
\begin{equation}\label{SCA38}
Q^1\,=\,\begin{pmatrix} Q^1_{\alpha}\\ 
Q^1_{\dot{\alpha}}\end{pmatrix};\qquad 
Q^2\,=\,\begin{pmatrix} Q^2_{\beta}\\ Q^2_{\dot{\beta}}\end{pmatrix}\;,
\end{equation}
where
\begin{eqnarray}\label{SCA39}
Q^1_{\alpha}&=& \oint\,d\,z\,
e^{-\,\vp_L/2}\,\sla\,\exp{\Big[\frac{i\,\sqrt{3}}{2}H_L\Big]}\,(z) \nonumber\\
Q_{1\,\dot{\alpha}}&=& \oint\,d\,z\,
e^{-\,\vp_L/2}\,\slb\,\exp{\Big[-\,\frac{i\,\sqrt{3}}{2}H_L\Big]}\, (z) \\
Q^2_{\beta}&=& \oint\,d\,\bz\,
\label{SCA40}
e^{-\,\vp_R/2}\,\sra\,
\exp{\Big[-\,\frac{i\,\sqrt{3}}{2}H_R\Big]}\,(\zbar) \nonumber \\
Q^{2\,\dot{\beta}}&=& 
\oint\,d\,\bz\,
e^{-\,\vp_R/2}\,\srb\,
\exp{\Big[\,\frac{i\,\sqrt{3}}{2}H_R\Big]}\,(\zbar) \nonumber
\end{eqnarray}
with $S_\alpha$ and $S_{\dot{\alpha}}$ as given above are the spin fields
of $SO(3,1)$ obtained by bosonising the fermions $\psi^\mu$ in the spacetime
sector.
The above choice reflect the spinorial content of the Ramond ground state
in ten-dimensions.
For the IIA string, the chiralities
are opposite in the left and right sectors. Let us choose them to be $\eights$
(i.e., positive ten-dimensional chirality)
for the left-movers and $\eightc$ for the right movers. 
From the Calabi-Yau sector, the chirality of the Ramond ground states is
reflected in sign of $U(1)$ charge. Since ten-dimensional chirality must
be given by the product of the four-dimensional and internal(CY) chirality,
one has: 
$\eights\rightarrow (\alpha,+)\oplus (\dot{\alpha},-)$ and
$\eightc\rightarrow (\alpha,-)\oplus (\dot{\alpha},+)$. The above choices
for the supersymmetry charges reflect this.

\subsubsection{Gravity multiplet and the universal 
hypermultiplet}\label{scasss2}

The vertex operators for the NSNS fields, i.e., the graviton, $B$-field and
the dilaton in the $(-1,-1)$ picture are:
\begin{equation}\label{SCA11}
V^{(-1, -1)}(k, \zeta) \,=\,
\zeta_{\mu\nu}\,e^{-\vp_L-\vp_R}\,\psi^{\mu}_L(z)\,
\psi^{\nu}_R(\zbar)\,e^{ik\cdot X}(z, \zbar)
\end{equation}
where $k^2 \,=\, k^{\mu}\,\zeta_{\mu\nu}\,=\,
k^{\nu}\,\zeta_{\mu\nu}\,=\,0$.
The {\em symmetric traceless} part of $\zeta_{\nu\mu}$ gives the graviton vertex
operator, the {\em antisymmetric} part, the vertex operator for the $B$-field 
and the {\em trace}, the dilaton vertex operator.

The states that come from the RR sector are the graviphoton:
($\vep_{\mu}$ be the corresponding polarization vector)
\ben\label{SCA25}
V^{(-\frac{1}{2}, -\frac{1}{2})}_{\rm{graviphoton}}(k, \vep) &=&
k_{[\,\nu}\,\vep_{\mu\,]}\,e^{-(\vp_L+\vp_R)/2}
\Big[\,\sla\,\epsilon^{\alpha\gamma}
(\sigma^{[\mu\nu]})_{\gamma}^{~\beta}\,\sra\,
\Sigma_0(z,\zbar)\non\\
&+&
\slb\,\epsilon_{\dot{\alpha}\dot{\gamma}}
(\ol{\sigma}^{[\mu\nu]})^{\dot{\gamma}}_{~\dot{\beta}}
\,\srb\,\Sigma^{0\dagger}(z,\zbar)\,\Big]\,\times\,e^{i\,k\cdot X}(z, \zbar)
\een

The RR scalars $\xi^0$ and $\xi_0$ that 
form the universal hypermultiplet (with the dualised $B$-field and the dilaton)
are given by the vertex operators:
\begin{equation}
V^{(-\frac{1}{2}, -\frac{1}{2})}_{\xi^0+i\xi_0}(k) =
k_{\mu}\,e^{-(\vp_L+\vp_R)/2}
\Big[\,\sla\,\epsilon^{\alpha\gamma}(\sigma^{\mu})_{\gamma\dot{\beta}}\,\srb\,
\Xi^{0}(z,\zbar)\Big]
\,e^{i\,k\cdot X}(z,\zbar)
\end{equation}

Note the appearance of the four-momentum $k$ in the RR vertex operators -- 
these reflect the fact that they couple to field strengths of $p$-form
gauge fields.

\subsubsection{Vertex operators for the hypermultiplets}\label{scasss4}

The vertex operators for the scalars $z^A$ that arise
from the NSNS sector are given by
\begin{equation}\label{SCA27}
V^{(-1,\,-1)}_{(z^{A})}(k)\,=\, 
e^{-(\vp_L+\vp_R)}\,\Pi^{A}(z, \zbar) \
e^{i\,k\cdot X}(z, \zbar)
\end{equation}

The scalars $\xi^A$ and $\xi_A$ from the RR sector arise from
\begin{equation}\label{SCA29}
V^{(-\frac{1}{2}, -\frac{1}{2})}_{(\xi^A+i \xi_A)}(k) =
k_{\mu}\,e^{-(\vp_L+\vp_R)/2}
\Big[\,\sla\,\epsilon^{\alpha\gamma}(\sigma^{\mu})_{\gamma\dot{\beta}}\,\srb\,
\Xi^{A}(z,\zbar)\Big]
\,e^{i\,k\cdot X}(z, \zbar)
\end{equation}

\subsubsection{Vertex operators for the vector multiplet}\label{scasss5}

The vertex operator for such scalars $t^j$ in ($-1$, $-1$)-picture is
given by :
\begin{equation}\label{SCA35}
V^{(-1,\,-1)}_{t^{j}}(k)\,=\,
e^{-(\vp_L+\vp_R)}\,\Lambda^{j}
(z, \zbar)\,
\,
e^{i\,k\cdot X}(z, \zbar)
\end{equation}
with the one for $\bar{t}^j$ obtained by the replacement
$\Lambda^{j}\rightarrow \Lambda^{j\dagger}$.
The vertex operator of the vector field in the multiplet
is given by
\ben\label{SCA37}
V^{(-\frac{1}{2}, -\frac{1}{2})}_{C^{j}_{\mu}}(k, \ol{\epsilon}) &=&
k_{[\,\nu}\,\ol{\varepsilon}_{\mu\,]}\,e^{-(\vp_L+\vp_R)/2}
\Big[\,\sla\,\epsilon^{\alpha\gamma}
(\sigma^{[\mu\nu]})_{\gamma}^{~\beta}\,\sra\,
\Sigma_j(z,\zbar)\non\\
&+&
\slb\,\epsilon_{\dot{\alpha}\dot{\gamma}}
(\ol{\sigma}^{[\mu\nu]})^{\dot{\gamma}}_{~\dot{\beta}}
\,\srb\,\Sigma^{j\dagger}(z,\zbar)\,\Big]\,\times\,e^{i\,k\cdot X}(z, \zbar)
\een

\subsection{The orientifold projection}\label{scass3}

Our discussion so far holds for  general anti-holomorphic involutions.
We will now specialise to the case where the anti-holomorphic involution
$\sigma$ is complex conjugation. A  naive guess for the
action of $\sigma$ on an NSNS field
$\Phi^{q_L\,q_R}_{h\,\bar{h}}$ is($q_L$, $q_R$ $>$ 0):\footnote{The case
when $q_L=q_R=0$ is more complicated. For the case of single minimal
models, for low $k$, see the discussion in ref. \cite{Blumenhagen:2001jb}.}
\begin{equation}\label{SCA46}
\sigma \;:\; \Phi^{\,q_L\,q_R}_{\,h\,\bh} \,\to\,
\Phi^{\,-\,q_L\,-\,q_R}_{\,h\,\bh}
\end{equation}
This can be seen to be consistent with the fact that in the LG model
corresponding to the Gepner model, all chiral superfields get mapped
to anti-chiral superfields.  
Of course, as we have seen, $\sigma$ in itself is not
a symmetry but the combination ${\cal O}\equiv \sigma\cdot\Omega$ is.
The action of $\Omega$ interchanges left and right movers. Thus,
using the above rule, it is easy to see that $(c,c)$ states get
mapped to $(a,a)$ states while $(a,c)$($(c,a)$) states get
mapped to $(a,c)$($(c,a)$) states. 

\subsubsection{Hypermultiplets}\label{scasss6}

It is thus not hard to see that for 
massless NSNS modes that arise from the $(c,c)$ and $(a,a)$ states,
that one linear combination of the two scalars that make up $z^A$
survives the orientifold projection. In the RR sector, one has
(for $A=0,1,\ldots,h_{1,2}$)
$$
\quad\sla \srb\ \Xi^A \stackrel{(-)^{F_L}\cdot\OO}{\longleftrightarrow}
(S_L)^{\dot{\beta}} (S_R)_\alpha \ \Xi^{A\dagger}
$$
and thus one linear combination of the scalars $\xi^A$ and $\xi_A$ survives
the projection. They combine to form a ${\cal N}=1$ chiral multiplet.

\subsubsection{Gravity and Vector multiplets}\label{scasss7}

Let us choose to write the identity operator as $\Lambda^0$.
The action of $\OO$ on the fields $\Lambda^{j}$ (choosing a diagonal
basis) is (no summation over $j$ below)
\begin{equation}\label{actransform}
\OO\, \Lambda^{j}\, \OO^{-1} = \nu_j\, \Lambda^{j}\ ,
\end{equation}
with $\nu_0=+1$ and $\nu_j=\pm1$ for $j=1,\ldots,h_{1,1}$. The spectral
flow operator $U_{\frac12,-\frac12}$ maps these operators to the
Ramond ground states $\Sigma^j$. $\OO$ has the following action on
$U_{\frac12,-\frac12}$:
\begin{equation}
\OO\, U_{\frac12,-\frac12}\, \OO^{-1} =\eta\, U_{\frac12,-\frac12}\ ,
\end{equation}
where we have included a possible sign. This implies that
\begin{eqnarray}\label{rgstransform}
\OO\, \Sigma^{j}\, \OO^{-1} = \eta\, \nu_j\, \Sigma^{j}\ .
\end{eqnarray}
Finally, we need the action of $\OO$ on the fermions $\psi_L^\mu$ and
$\psi_R^\mu$:
\begin{equation}\label{gravtransform}
\OO\, \psi_L^\mu\,\psi_R^\nu\, \OO^{-1} = \psi_L^\nu\,\psi_R^\mu
\end{equation}

\noindent

Using eqn. (\ref{gravtransform}), one can see that the symmetric
part of the first vertex operator is projected in 
implying that the graviton and dilaton are projected in. Now, 
Using equation (\ref{actransform}), we can see that when $\nu_j=+1$,
the scalars that come from the second vertex operator above, is
projected in.

In order to work out the orientifold projection, we need to consider
the action of $\OO$ and $(-)^{F_L}$ on the spinfields.
\begin{equation}
\OO \sla\,
\sra\ \OO^{\,\dag} = (S_R)_\alpha (S_L)_\beta = -(S_L)_\beta \, (S_R)_\alpha
\end{equation}
where the minus sign comes from the cocycles (see for instance, 
\cite{CALLAN,GP})
 or equivalently from the fact that
the spin fields are spacetime fermions. Under $(-)^{F_L}$, one has
\begin{equation}
(S_R)_\alpha (S_L)_\beta \xrightarrow{(-)^{F_L}} -(S_L)_\alpha \, (S_R)_\beta
\end{equation}
Thus, under the combined action of $\OO$  and $(-)^{F_L}$, one obtains
\begin{equation}
(S_R)_\alpha\, (S_L)_\beta\, \Sigma^j \xrightarrow{\OO\cdot(-)^{F_L}}
\eta\, \nu_j\, (S_L)_\beta \, (S_R)_\alpha\, \Sigma^j
\end{equation}
Thus, for $\eta\nu_j=+1$, the symmetric part is projected in. One can
show that $\epsilon^{\alpha\gamma} (\sigma^{[\mu\nu]})_{\gamma}^{~\beta}$
is a symmetric matrix(see Appendix A, \cite{WB})
and hence the gauge field is projected in. In particular,
we would like to see that the graviphoton has to be projected out, i.e.,
we need $\eta\nu_0=\eta = -1$. Thus, we need $\eta=-1$. Once this choice
is made, we see that, for $\nu_j=-1$, the vector multiplet is projected
in. Here is a summary of states projected in given in table 2.
\begin{table}[!ht]
\begin{center}
\begin{tabular}{|c|c|c|}\hline
Sector & NSNS & RR \\ \hline
Graviton & Symmetric part & -- \\\hline
$\nu_j=+1$ & scalars & -- \\\hline
$\nu_j=-1$ & -- & vector \\ \hline
\end{tabular}
\caption{Summary of the projection in gravity and vector multiplets} 
\end{center}
\end{table}
Comparing with the M-theory analysis, we see that $\nu_j=-1$ corresponds
to the K\"ahler class be even. These results are consistent with the
geometric analysis of ref. \cite{Vafa:1995gm}.

\sectiono{Basics of Unoriented Strings}\label{bs2}

In this section we summarize the main results of \cite{CARDY} and 
\cite{PSS} for constructing RCFTs on unoriented worldsheets with and 
without boundary.
At the level of CFT's, an 
orientifold introduces surfaces with crosscaps, \ie
unoriented string sectors\fn{It may or may not introduce (unoriented) 
open string sectors, \ie D-branes. It very much depend on the nature of 
the orientifold.}.  
We start with a known modular invariant torus 
partition function for oriented closed string sector
\begin{eqnarray}\label{b2.1}
\TO &=& 
\sum_{i\,,\,j}\,\chi_i\,Z_{i\,j}\,\ol{\chi_j}\,,\\
\text{with}\quad \chi_i(\tau) &=& {\rm Tr}_{\HH_i}\,q^{L_0 -
c/24}\,,\;\; q \,=\, e^{2\pi\,i\,\tau}
\end{eqnarray}
where $Z_{i\,j}$ is a symmetric modular invariant matrix and $\chi_i$ is 
the character of the representation $i$;
here $\HH_i$ is the Hilbert space for the $i$-th irreducible 
representation of the algbera.

\medskip

\noindent$\bullet\;\;$ {\bf Klein bottle:}
\smallskip

Suppose the orientifold group is\fn{$\OO$ may or may not contain $\wil$ 
factor.} $\OO = \Omega\cdot\sigma$, where $\sigma$
is an anti-holomorphic involution of the $\CY_3$; in CFT language it
corresponds to {\em simple currents} of the underying worldsheet algebra.
Since crosscap interchanges chiral with anti-chiral fields of 
the 2d RCFT, therefore\footnote{For notation and conventions, see appendix 
\ref{A1}.},
\begin{equation}\label{b2.5}
\OO\;:\; \varphi_{\,i\,i^c} \,\to\, \ep_i\,\vp_{\,i^c\,i}\,;\quad 
\OO^2 \,=\, \one\quad\ep_i=\pm 1\,,
\end{equation}
where \{$\varphi_{\,i\,i^c}$\} denote the set of primary fields of the 
RCFT, 
labelled by their conformal weights ($h_i$, $\bh_i$). 
The {\em direct channel} Klein bottle amplitude  is defined 
as follows:
\ben
\KB_{\text{NSNS}\atop \text{RR}}(q) 
&=& \half\,{\rm Tr}_{\text{NSNS}\atop\text{RR}}
\big[\,\Omega\cdot\sigma\,(1\,+\,\wi\,\,)q^{H_{cl}}\,\big]\non\\
&=& 
\frac{1}{2}\,\sum_{i}\,\KB^i\,
\chi_i^{\text{(NSNS)}\atop\text{(RR)}}(q)\,,
\label{b2.6a}
\een
where $F$ is the worldsheet fermion number operator, $\KB^i$ are 
integers 
 and $H_{cl} = 
\frac{1}{2}\,\big(\,L_0 \,+\,\Lbar_0 \,-\, \frac{c}{12}\,\big)$ is the 
closed string hamiltonian. 
The condition that the closed string sector 
$\dfrac{1}{2}\,\big(\,\TO\,+\,\KB\,\big)$ has positive, integral 
multiplicities for all states requires that\cite{PSS}
\begin{equation}\label{b2.7}
\KB_i = \ep_{\,i}\,Z_{i\,i}\,,
\end{equation}
where we define $\ep_i=0$ when $Z_{i\,i}=0$.\fn{Eqn. \refb{b2.7} is 
not all. It is consistent (under the operator product expansion)
iff $\ep_i\,\ep_j\,\ep_k\,\NN_{i\,j}^{\;\;k}\,\ge\,0$,
where $\NN_{i\,j}^{\;\;k}$ are the fusion matrix elements, given by the 
Verlinde formula\cite{VERLINDE} : 
$\NN_{i\,j}^{\;\;k}\,=\,\sum_l\,\frac{S_{i\,l}\,S_{j\,l}\,
S_{k\,l}^{\dag}}{S_{0\,l}}$.
}
In the {\em transverse channel} Klein bottle amplitude
is interpreted 
as the propagation of closed strings between two crosscap states :
\begin{equation}\label{b2.10}
\wt{\KB}(\tq) = \Bra{C}\,\tq^{H_{cl}}\,\Ket{C}\,=\,
\sum_{i}\,\Gamma^i\,\Gamma^i\,\chi_i(\tq)\,,
\end{equation}
where in the last step we have used eqn. \refb{A.4} and $\tq \,=\, 
e^{-\,2\pi\,i/\tau}$.
Since eqns.\refb{b2.6a} and \refb{b2.10} are related by modular 
transformation matrix $S$, 
we have the following (consistency) condition
\begin{equation}\label{b2.11}
\KB_i= \sum_{j}\,S_i\,^j\,\Gamma_j\,\Gamma_j
\end{equation}

\medskip

\noindent$\bullet\;\;$ {\bf M\"obius strip :}
\smallskip

Generically, the unoriented closed string theory is inconsistent due to 
the presence of massless tadpoles. One can get rid of them by introducing 
open string sectors, \ie, D-branes. 
At the worldsheet level, the relevant 1-loop open string amplitudes which 
contain these open string sectors are the annulus and the M\"obius strip. 
The M\"obius strip amplitudes are defined in terms of real characters:
\begin{equation}\label{b2.15}
\wh{\chi_i} = 
e^{i\,\pi\,(h_i\,-\,c/24)}\,\chi_i\Big(\,\frac{i\,\tau\,+\,1}{2}\,\Big)
= 
\big(\,\sqrt{T}\,\big)^{-1}\,\chi_i\,\Big(\,\frac{i\,\tau\,+\,1}{2}\,\Big)
\end{equation}
The modular transformation matrix connecting the direct and transverse 
channels for the M\"obius amplitudes is\fn{Like $S$, $P$ is a unitary and 
symmetric matrix and satisfies
\begin{equation}\label{b2.17}
P^2=\CC\,,\quad P^{\ast}\,=\,\CC\,P\,=\,P\,\CC\,.
\end{equation}
}
 given by\cite{PS} :
\begin{equation}\label{b2.16}
P= \sqrt{T}\,S\,T^2\,S\,\sqrt{T}\,,
\end{equation}
The direct and transverse channel M\"obius strip amplitudes are
\begin{eqnarray}\label{b2.18}
\text{Direct}\quad:\quad
\MS_{\text{NS}_a\atop\text{R}_a}(\tau) 
&=& {\rm 
Tr}_{\text{NS}_a\atop \text{R}_a}\,
\big(\,\Omega\,h\,q^{L_0\,-\,c/24}\,\big)\,=\,
\sum_i\,\MS^i_{\text{NS}_a\atop \text{R}_a}\,
\wh{\chi}^{\text{NS}_a\atop 
\text{R}_a}_i(\tau)\,,\non\\
\text{Transverse}\quad :\quad\wt{\MS}_{\text{NS}_a\atop 
\text{R}_a} 
&=&
\Bra{C}\,\tq^{H_{cl}}\,\Ket{B,\dps\text{NSNS}_a\atop \text{RR}_a}\,=\,
\sum_i\,\Gamma^i\,B^i_{\text{NSNS}_a\atop 
\text{RR}_a}\,
\wh{\chi}_i^{\text{NS}_a\atop 
\text{R}_a}(\tq)\,,
\end{eqnarray}
where $\MS^i_{\text{NS}_a\atop \text{R}_a}$ 
represents twists of open string spectra in NS- and R-sectors 
respectively; these are 
non-negative integers and $a$ specifies the boundary condition on the 
boundary. Also $\Ket{B\,;\,{\text{NSNS}_a\atop 
\text{RR}_a}}$ denotes either NSNS- or RR- boundary state for 
a 
D-brane satisfying a set of  boundary conditions $a$. As in case of Klein 
bottle, the total M\"obius strip amplitude is obtained by asumming the 
contribution of NS- and R-sectors of open string channel. 
The channel transformation matrix $P$ relates $\MS_a(\tau)$ and 
$\wt{\MS_a}$ as follows :
\begin{equation}\label{b2.20}
\MS^i_{\;a}\,=\,\sum_j\,\Gamma^j\,B^j_{\;a}\,P_j^{\;i}\,,
\quad\forall\quad i,\,a\,.
\end{equation}

\subsection{Techniques of open descendants}\label{bss2.1}

Given a modular 
invariant, $Z_{i\,j^c}$ and a consistent Klein bottle 
projection, $\KB_i$ in the closed string sector of a RCFT, subject to 
the constraints given in the appendix \ref{A2}, the answer to
the correct open string spectra or in other word, the 
correct annulus and M\"obius coefficients, $\AN^{\,i}_{\;\;a\,b}$, 
$\MS^{\,i}_{\;\;a}$ 
has been given in ref.\cite{PSS}. Henceforth, these solutions
will be referred to as PSS solutions. We just quote their results 
here only for Klein bottles and
M\"obius strips. There are two types of solutions; the first one being 
the well-known Cardy type\cite{CARDY}

\noindent{\bf Cardy-type solutions:}

We start with a $\CC$-diagonal modular invariant, \viz, 
$Z_{i\,j}\,=\,\delta_{i\,j^c}$. The values of Klein bottle and M\"obius 
strip 
coefficients for the  $\CC$-diagonal case can 
be expressed in terms of an {\bf integer-valued tensor}(either 
+ve or $-$ve)\cite{PSS} :\fn{Defining the matrix $Y_i$ as 
$\big(\,Y_i\,\big)_j^{\;k}=
Y_{\,i\,j}^{\;\;k}$,
one can check that
these matrices $Y_i$ are mutually commuting and satisfy the fusion 
algebra : $Y_i\,Y_j\,=\,\sum_k\,\NN_{\,i\,j}^{\;\;k}\;Y_k$.}
\begin{equation}\label{b2.23}
Y_{\,i\,j}^{\;\;k}\,=\,\sum_m\,
\frac{S_{i\,m}\,P_{j\,m}\,P_{k\,m}^{\,\ast}}{S_{0\,m}}
\end{equation}
In terms of 
$Y_{\,i\,j}^{\;\;k}$, the M\"obius, Klein Bottle amplitudes, boundary and 
crosscap coefficients are :
\begin{eqnarray}
\text{{\bf M\"obius }} \quad:\quad 
\MS^{\,i}_{\;a} &=& Y_{a\,0}^{\;\;i}\label{b2.26}\\
\text{{\bf Klein Bottle }}\quad:\quad
\KB^{\,i} &=& Y^i_{\,\;0\,0}\label{b2.27}\\
\text{{\bf Crosscap coefficients }} \quad:\quad
\Gamma_i &=& \frac{P_{i\,0}}{\sqrt{S_{i\,0}}}\quad\imply\;\;
\boxed{\ket{C}\,=\,
\sum_i\,\frac{P_{i\,0}}{\sqrt{S_{i\,0}}}\,\ket{C_i}}\label{b2.29}.
\end{eqnarray}
Eqn. \refb{b2.29} is the basic building block for 
constructing crosscap states in Gepner model 
and will be used in sections \ref{ns1} and \ref{qs1} for constructing 
crosscaps in $(k=1)^3$ and $(k=3)^5$ Gepner models.

\subsection{Order $\mathbf{\N}$ simple currents and
corresponding open descendants}\label{bss2.2}

The other solutions to the 
algebraic constraints of appendix \ref{A2} 
are given in terms of $S$, $P$ and $Y$
matrix elements associated to certain primary fields, called {\em simple
currents}\cite{SY}. These extra solutions of those constraints in terms of
simple currents are relevant for our purpose because the antiholomorphic
involutions, generically denoted as $\sigma$ in the orientifold group 
is in one-one correspondence with simple currents of the RCFT. These 
solutions are given in ref.\cite{HSS}. 
\begin{eqnarray}
\text{{\bf Klein Bottle}} \quad:\quad
\KB^i_{\,[L]} 
&=& \,e^{2\,\pi\,i\,Q_{L^n}(\Phi_{\,i})}\,Y^i_{\,\;0\,0}\,=\, 
Y^i_{\,\;L\,L}\,,\label{b2.30} \\
\text{{\bf M\"obius Strip }}\quad :\quad
\MS^i_{\;[L^n]\,a}&=& \YY^i_{\,\;a\,L}\quad
\Big|\,\MS^i_{\;[L^n]\,a}\,\Big| \;\le\;
\NN^{\,L^n\,\ox\,\Phi_{\,i}}_{\;\;\;a\,a}\non\\
\MS^i_{\;[L^n]\,a} &=& \NN^{\,L^n\,\ox\,\Phi_{\,i}}_{\;\;\;a\,a} \mod 2\,,
\quad\forall\;n\,=\,0,\,1,\,\cdots\,,\N\,-\,1\non\\
& & \label{b2.31}\\
\text{{\bf Crosscap coefficients }}\quad :\quad
\Gamma_{\,[L^n]\,i} &=& \frac{P_{L^n\,i}}{\sqrt{S_{0\,i}}}\,,
\quad\forall\;n\,=\,0,\,1,\,\cdots\,,\N\,-\,1\label{b2.35}\\
\imply\;\;\ket{C\,;\,[L^n]} &=&
\sum_i\,\frac{P_{L^n\,i}}{\sqrt{S_{0\,i}}}\,\ket{C\,;\,i}\,,\label{b2.36}
\end{eqnarray}
where $Q_{L^n}(\Phi_{\,i})$ is the monodromy charge of the primary 
field
$\Phi_{\,i}$, as given by eqn. \refb{C.5}. 
Like eqn. \refb{b2.29}, eqn. \refb{b2.36} 
will be the basic building block for constructing crosscap 
states associated to some particular simple currents in the Gepner model.

\sectiono{Crosscap states for Gepner models}\label{cs1}
\subsection{Boundary and Crosscap States in the minimal model}\label{css1}

The A-type boundary and crosscap states (associated with the simple current
$(0,M,S)$)  in the minimal model at level $k$ are
given by a straightforward application of the Cardy and PSS ansatz
to be\fn{Since we shall exclusively deal with A-type crosscap states, 
henceforth we drop the subscript $A$.}
\begin{eqnarray}\label{c1}
|B:L,M,S\rangle &\equiv& \sum_{(l,m,s)}^{{\sf FR}} 
\left(S_{(0,0,0)~(l,m,s)}\right)^{-\frac12 }\ S_{(L,M,S)~(l,m,s)} \
|B:l,m,s\rangle\rangle \\
|C:M,S\rangle &\equiv& \sum_{(l,m,s)}^{ {\sf FR}}
\left(S_{(0,0,0)~(l,m,s)}\right)^{- \frac12 }\ P_{(L,M,S)~(l,m,s)} \
|C:l,m,s\rangle\rangle 
\end{eqnarray}
The $S$ and $P$-matrices for the minimal model case have been discussed
in the appendix \ref{A4s3}.
Putting in the values of the $S$ and $P$-matrices, an explicit expression
for the the boundary and crosscap states are(with $(L+M+S)=$even for the 
boundary
states and $(M+S)=$even for the crosscap states):\footnote{For the choice of 
$\wh{\sigma}$ given in eqn. (\ref{sigmadef}), the crosscap 
state is identical to the one constructed by
Brunner and Hori\cite{BH2}.}
\begin{equation}\label{c2}
\begin{aligned}
|B:L,M,S\rangle &\equiv \sqrt{\frac1{\sqrt2(k+2)}} 
\sum_{(l,m,s)}^{ {\sf FR}} 
\frac{\sin(L,l)_k}{\sqrt{\sin(l,0)_k}}
\exp\left(\frac{i\pi Mm}{k+2}
- \frac{i \pi Ss}{2}\right)\ 
|B:l,m,s\rangle\rangle \\
|C:M,S\rangle &= \sqrt{\frac2{k+2}} \sum_{(l,m,s)}^{{\sf FR}}
\frac{ \sigma_{l,m,s}}{\sqrt{\sin(l,0)_k}}  \
\exp i \pi\left(\frac{Mm}{2k+4} - \frac{Ss}4\right)\ 
\delta^{(2)}_{s+S}\times \\
& \Big[\sigma_{0,M,S}\  \sin\frac12(0,l)_k\
\delta^{(2)}_{l+k}\ \delta^{(2)}_{M+m+k} \hspace*{3in}  \\
&+ \sigma_{k,M+k+2,S+2}   \
(-)^{\frac{-l+m-s}2}\
\cos\frac12(0,l)_k\
\delta^{(2)}_{l}\ \delta^{(2)}_{M+m}\Big]\Big|C:l,m,s\Big\rangle\Big\rangle  
\end{aligned}
\end{equation}
where $\big(l,l'\big)_k=\pi(l+1)(l'+1)/(k+2)$ and
$\sigma_{l,m,s}\equiv \widehat{\sigma}_{l,m,s}\nu_m^{(k+2)} \nu_s^{(2)}$.
($\widehat{\sigma}_{l,m,s}$ is defined in eqn. (\ref{sigmadef}) and
$\nu_m^{(k)}$ is defined in eqn. (\ref{nudef}) in Appendix \ref{A4}.)
We have also used the identity
$$\delta^{(2)}_l\sin\frac12(k,l)_k = \delta^{(2)}_l
(-)^{l/2}\cos\frac12(0,l)_k\ .$$ 
It is important to note that when $S$ is even, the crosscap states consist
of Isibashi states exclusively from the NSNS sector and when $S$ is odd,
the crosscap states consist of Isibashi states exclusively from the RR sector.

\noindent Under the various identifications, one can verify that
the boundary and crosscap states transform as:
\begin{equation}\label{c3}
\begin{split}
\big|B:L,M,S\big\rangle = \big|B:L,M+2k+4,S\big\rangle= 
\big|B:M,S+4\big\rangle= \big|B:k-L,M+k+2,S+2\big\rangle \\
\big|C:M,S+4\big\rangle=(-)^S\ \big|C:M,S\big\rangle
\end{split}
\end{equation}
The second line above follows on using
$\sigma_{l,m+2k+4,s}=\sigma_{l,m,s+4}=\sigma_{l,m,s}$. 
The Klein bottle amplitudes for the above crosscap states are\cite{BH2}
\begin{equation}\label{kbmm}
\boxed{
\begin{split}
\langle C:\tilde{M},\tilde{S}|\tilde{q}^{H_cl}|C:M,S\rangle =
\sigma_{0,\tilde{M},\tilde{S}}\ \sigma_{0,M,S}\ \delta^{(2)}_{M+\tilde{M}}\
\delta^{(2)}_{S+\tilde{S}}\hspace*{2.6in}\\
\Big[\sum_{l}^{ev} (-)^l \Big(
\chi_{l,\frac{\tilde{M}-M}2}^{(\frac{\tilde{S}-S}2)}(q) 
+ (-)^S \chi_{l,\frac{\tilde{M}-M}2}^{(\frac{\tilde{S}-S}2+2)}(q)\Big)
+\delta^{(2)}_{k} \Big(\chi_{\frac{k}2,\frac{\tilde{M}-M+k+2}2}^{(\frac{\tilde{S}-S+2}2)}(q) 
+ (-)^S \chi_{\frac{k}2,\frac{\tilde{M}-M+k+2}2}^{(\frac{\tilde{S}-S-2}2)}(q)\Big)\Big]
\end{split}
}
\end{equation}
It is useful to observe that (for odd $k$)
when $S-\tilde{S}=0\mod4$, the  characters
$NS\equiv \text{Tr}_{NS}\big[\Omega\, q^{L_0}\big]$
appear for even $S$ and the  characters 
$R\equiv \text{Tr}_{R}\big[\Omega\, q^{L_0}\big]$
appear for $S-\tilde{S}=2\mod4$. For odd $S$,
the characters with the $(-)^F$ insertions appear,
$\widetilde{NS}\equiv \text{Tr}_{NS}\big[\Omega\, (-)^F\,q^{L_0}\big]$ when
$S-\tilde{S}=0\mod4$  and 
$\widetilde{R}\equiv \text{Tr}_{R}\big[\Omega\, (-)^F\,q^{L_0}\big]$ when 
$S-\tilde{S}=2\mod4$.\fn{Here we are somewhat sloppy about our notation; 
the full orientifold group also includes the generator for antiholomorphic
involution, the generators of discrete symmetry group of the Gepner model 
and $\wil$ -- the presence of the latter being depended on the 
dimension.}
When $k$ is even, the second term contributes in an opposite fashion,
i.e., $R$ appears for even $S$ and $S-\tilde{S}=0\mod4$ and so on.
For the boundary states, a similar role is played by the 
combination\cite{SGTJLG}
\begin{equation} \label{fullbstate}
|B:L,M,\pm\rangle \equiv \frac1{\sqrt2}\Big(|B:L,M,S\rangle \pm
|B:L,M,S+2\rangle\Big)\quad, 
\end{equation}
One can easily see that
$|B:L,M,+\rangle$ involve Ishibashi states from the NSNS sector and
$|B:L,M,-\rangle$ involve Ishibashi states from the RR sector. 
The annulus channel among the overlap of the $|B:L,M,+\rangle$ boundary states
involves $NS$ (for even $(L+M)$  and $R$ amplitudes (for odd $(L+M)$)
and the $\widetilde{NS}$ and $\widetilde{R}$ characters appear in the annulus
channel of the overlap 
between $|B:L,M,+\rangle$ and the $|B:L',M',-\rangle$. 
These will be the building blocks for the Gepner model boundary and
crosscap states.

\subsection{Discrete symmetries in the minimal model}\label{css2}

The minimal models have discrete symmetries generated by the simple
currents  with labels $(0,m,s)$. For odd $k$, the group is
$\BZ_{4k+8}$ (generated by the simple current given by the primary
with labels $(0,1,1)$) and for even $k$, the group is $\BZ_{2k+2}\times
\BZ_2$ (generated by the primaries $(0,1,1)$ and $(0,0,2)$). We focus
on three generators, which we will call $g$ (corresponds to $(0,2,0)$,
$h$ (corresponds to $(0,0,2)$) and $f$ (corresponds to $(0,1,1)$ and
generates spectral flow with $\eta=1/2$). Note the identities 
\begin{equation}
g\cdot h=f^2\ ,
\ h^2=1\  \text{and } g^{k+2}=1\ . 
\end{equation}
However, these identities may be projectively
realised on crosscap states. For instance, 
\begin{equation}\label{symmcs}
h^{2}\ |C:M,S\rangle = (-)^S \ |C:M,S\rangle\ .
\end{equation}
Thus, one will need to add phases into the action of $g$  and $h$ to get 
their $(k+2)$-th and second powers respectively to equal one on crosscap
states.

Under the discrete symmetries of the
minimal model generated by $g$ and $h$,  the boundary and crosscap
states transform as
\begin{eqnarray}
g\cdot |B:L,M,\pm\rangle &=& |B:L,M+2,\pm\rangle \quad,\quad
h\cdot|B:L,M,\pm\rangle =\pm |B:L,M,\pm\rangle \\
g\cdot |C:M,S\rangle &=& |C:M+2,S\rangle \quad,\quad
h\cdot|C:M,S\rangle =\ |C:M,S+2\rangle
\end{eqnarray}
The boundary and crosscap states form orbits of length $(k+2)$
with one exception.  When $k$ is even and
$L=k/2$, the orbit length is $n=(k+2)/2$.
$$
g^{n}\cdot |B:L,M,\pm\rangle = \pm |B:L,M,\pm\rangle\quad.
$$
In the context of the Gepner model, this leads to the
boundary states which are not minimal and need to be resolved.

\subsection{Discrete Automorphism Group of $(k=1)^3$
and $(k=3)^5$ Gepner Models}
\label{nsss1.2}
\medskip

Once we know the discrete symmetries of each minimal model, we can 
specify
the orientifold group of a Gepner model. This group is model as
well as theory(\ie whether type IIA or IIB) dependent. 
As we saw in sections \ref{bs4} and \ref{bss4.1},
generically for type IIA compactification down to four spacetime 
dimensions, 
this group is given by\fn{The subscript $k$ in the definition of $G$ 
reminds us of the level the minimal models used for compactification. The 
group $G$ depends crucially on it.} $G_k\,=\,
\Omega\cdot\sigma\cdot\HH\cdot\wil$, where $\HH$ is some subgroup of the 
discrete automorphism group
of the Gepner models associated to the Calabi-Yau hypersurface. As we
 shall be dealing with Gepner models with all $k_i = \text{odd}$, from 
the
discussion of the preceding section it is quite clear that the 
automorphism
group is $\GG\sim \prod_i \BZ_{k_i+2}$ which is generated by 
$\prod_ig_i$. The
group element acts on the NSNS sector stated by 
multiplying them by 
$e^{\frac{2\pi i t_i(q_{L,i}\,+\,q_{R,i})}{2(k_i+2)}}$ with $t_i\in \BZ$.
\fn{For $A$, $D_{\text{odd}}$ and $E_6$ models, $t_i \in \BZ$ and $t_i\in
2\BZ$ for $D_{\text{even}}$, $E_7$ and $E_8$ models.}
The transformation corresponding to 
$t_i=(2,2,\cdots,2)$ is a trivial one and hence we mod out by the center
$\BZ_{\tn}\subset\prod_r\,\BZ_{n_i}$, where $\tn = \text{lcm}\;(k_i+2)$.
\fn{If not all $k_i$ are {\it odd}, $\tn = \half\text{lcm}\;(k_i+2)$.} 
Further, we can consider the permutation group which acts by permuting 
$r$
copies of minimal models amongst themselves. So\cite{FUCHSETAL}
\fn{There is an subtlety in defining $n_r$. It depends on which type
of modular invariants are being used to build up the torus parition 
function
of the theory. For $A$, $D_{\text{odd}}$ and $E_6$ modular invariants, 
$n_i = k_i+2$. For $D_{\text{even}}$, $E_7$ and $E_8$ modular 
invariants, we have  $n_i=(k_i + 2)/2$. In our examples of $(k=1)^3$ and 
$(k=3)^5$ Gepner models, it is always given by $(k_i+2)$.}
$$
\GG = \frac{\prod_i\BZ_{n_i}}{\BZ_{\tn}}\ox S_r = 
\frac{\prod_i\BZ_{k_i+2}}{\BZ_{\tn}}\ox S_r
$$
We do not consider the permutation group and instead consider the 
subgroup
$\HH = \frac{\prod_r \BZ_{k_r+2}}{\BZ_{\tn}}$. For $(k=1)^3$ model,
$\HH = \BZ_3^3/\BZ_3$ and for $(k=3)^5$ model, it is 
$\HH = \BZ_5^5/\BZ_5$. So the orientifold group for these models are
\fn{Since compactification of type 
IIA theory on $(k=1)^3$ model($c=3$) corresponds to compactification down 
to 8 
spacetime-dimension, the orientifold group does not involve $\wil$; see 
eqn.\refb{b2.48a}.}
\begin{equation}\label{og}
G_{(k=1)^3}\,=\,\Omega\cdot\sigma\cdot\frac{\BZ^3_3}{\BZ_3}\,;\quad
G_{(k=3)^5}\,=\,\Omega\cdot\sigma\cdot\wil\cdot\frac{\BZ^5_5}{\BZ_5}
\end{equation}

\subsection{The spacetime sector}\label{css3}

We will base our discussion here in the light-cone gauge. So if the
non-compact spacetime is $D$ dimensional, then $2d=D-2$. So for the
$1^3$ Gepner model, $d=3$ and $3^5$ Gepner model, $d=1$. Thus, we will focus
on the case when $d$ is odd in our examples.
The four irreps of $SO(2d)_1$ are the
scalar, vector, spinor and conjugate spinor ($O,V,S,C$) representations.
The weights and charges are $[(0,0),(1/2,\pm1),(d/8,d/2),(d/8,-d/2)]$ 
respectively.
We will represent them by the labels $s_0=0,2,1,-1$. Their $U(1)$ charges
are given by $(ds_0/2)\mod 2$.
The boundary and crosscap states (for $SO(2)$ and $SO(6)$) are:
\begin{equation}\label{spacetimebc}
\begin{split}
\Big|B_{\rm st}:S_0\Big\rangle &= \frac{1}{\sqrt2} \sum_{s_0} 
\exp\Big(-\frac{id\pi s_0 S_0}2\Big)
\Big|B_{\rm st}:s_0\Big\rangle\Big\rangle \\
\Big|C_{\rm st}:S_0=0,2\Big\rangle &= (-1)^{\frac{(d-1)}2\frac{S_0}2}\  
\sum_{s_0} 
\exp\Big(-\frac{i\pi s_0 S_0}4 \Big) \Big[
\cos(d\pi/4)\ \delta^{(4)}_{S_0-s_0} +
\sin(d\pi/4)\ \delta^{(4)}_{S_0-s_0+2}\Big] \Big|C:s_0\Big\rangle\Big\rangle\\
\Big|C_{\rm st}:S_0=\pm1\Big\rangle &=  (-1)^{\frac{(d-1)}2\frac{S_0}2}\ 
\sum_{s_0} 
\exp\Big(-\frac{id\pi}4 \Big) \Big[\cos(d\pi/4) \ \delta^{(4)}_{S_0-s_0}+
i\  \sin(d\pi/4)\ \delta^{(4)}_{S_0-s_0+2}\Big] \Big|C:s_0\Big\rangle\Big\rangle
\end{split}
\end{equation}
We introduce the phase factor in the definition of the crosscap above, to 
take care of the correct relative sign between the NSNS and RR parts of 
the KB amplitude for the $(k=1)^3$ Gepner model(here $d=3$). It has no 
effect for the $(k=3)^5$ model(here $d=1$).

The Klein bottle amplitude that is obtained from the above crosscap
states are given by (for odd $d$)
\begin{equation}\label{externalkb}
\begin{aligned}
\langle C_{\rm st}:\tilde{S}_0|e^{\frac{-i\pi}{2\tau}H_{cl}}|C_{\rm 
st}:S_0\rangle
&= (-1)^{\frac{(d-1)}2\frac{(S_0 - \wt{S}_0)}2}\ 
\frac12\big[\delta_{S_0-\tilde{S}_0}^4  \big(\chi_{\rm 
st}^{(0)} +(-)^{S_0}
\chi_{\rm
st}^{(2)}\big)(2\tau)
+(-)^{S_0}\delta_{S_0-\tilde{S}_0+2}^4\\
&\times  \sin(d\pi/2) \big(\chi_{\rm 
st}^{(1)}
+(-)^{S_0} \chi_{\rm st}^{(-1)}\big)(2\tau)\big]
\end{aligned}
\end{equation}
where $\chi_{\rm st}^{S_0}$ are the four $SO(d)_1$ characters. Note
that the spacetime KB amplitudes are quite similar to those that
we obtained for odd $k_i$ minimal models. In particular, the
discussion after eqn. (\ref{kbmm}) holds here as well. For later
 convenience, we define
 \begin{equation}\label{extchi}
 \chi_{\rm st}^{\pm,(NS)}\equiv(\chi_{\rm st}^{(0)}\pm \chi_{\rm st}^{(2)})
 \quad\text{and}\quad
 \chi_{\rm st}^{\pm,(R)}\equiv (\chi_{\rm st}^{(1)}\pm\chi_{\rm st}^{(-1)})
 \end{equation}

\subsection{The Gepner model}\label{css4}

Let us now consider the case of the Gepner model where the internal
CFT (corresponding to the Calabi-Yau manifold) is given by a Gepner
model, i.e., we consider the tensor product of $r$ 
minimal models of levels $k_i$ ($i=1,\ldots,r)$. The Gepner projection 
consists of the following:
\begin{enumerate}
\item Tensor NS states with NS states and R states with R states.
At the level of partition functions, it must consist of $NS$ characters
tensored with other $NS$ characters with similar conditions for
the other three types of characters, $\widetilde{NS}$, $R$ and $\widetilde{R}$.
We will call this the $\beta_r$-projection.
\item Project onto states such that the total $U(1)$ charge is an odd
integer. The total $U(1)$ charge has two contributions, one from the
spacetime CFT and one from the internal CFT, i.e., the Gepner model.
We call this the $\beta_0$ projection.
This can be done in a two-step process. For the internal CFT, first
consider states such that the total $U(1)$ charge is an integer
in the NS sector and (for odd $d$) half-integer in the R-sector.
Second, tensor the spacetime part in such a way that the total 
$U(1)$ charge is an odd integer. So if one has a state obtained from
the internal NS sector that has odd(even) $U(1)$ charge, then tensor it
with the NS representation of $SO(2d)_1$  with even charge, i.e.,
the scalar(vector) representation. 
\end{enumerate}

\subsubsection{The $\beta_r$-projection}\label{csss4.1}

As 
we saw in the discussion of a single minimal model as well as the
spacetime sector, the $S$ even crosscap states contain only $NSNS$ Ishibashi 
states and $S$ odd, $RR$ Ishibashi states. Thus, the first part of the
$\beta_r$-projection translates into the rule that one must tensor
states with even $S$ or odd $S$. From the Klein bottle amplitude
of the individual minimal models, eqn. (\ref{kbmm}), for odd $k$,
we see that we should impose something a little more stringent, i.e.,
require that we tensor all states with identical $S$.
Thus the following
tensor product of states (we assume that the level $k$ is odd
for all minimal models)
\begin{equation}\label{totalcrosscap}
|C:\mu,S\rangle \equiv |C_{\rm st}:S\rangle\ \prod_{i=1}^r |C:M_i,S\rangle \ .
\end{equation}
(with $\mu\equiv(M_1,\ldots,M_r)$)
has the following property that the loop channel of the overlap
of the crosscap does {\em not} mix different kinds of 
characters.  This clearly implements the
$\beta_r$-projection  when the $k_i$ are all odd. This will 
presumably require some
modification when even $k_i$ are involved.

\subsubsection{The  $\beta_0$-projection}\label{csss4.2}

We still have to implement the projection onto states with integer
$U(1)$ charge.   This is equivalent to orbifolding by the discrete group 
$\BZ_{K}$ ($K={\rm lcm}[4,2(k_i+2)]$) generated by 
$\widehat{g}\equiv [h_0g_1h_1\ldots g_rh_r]$. This symmetry is generated by
simple current obtained by tensoring the $(0,2,2)$ fields in the individual
minimal models.

The simple way to obtain crosscap states in the orbifold from the unorbifolded
theory is to consider the linear combination of all crosscaps that are
in a single orbit of the orbifolding group. We naively expect something
of the form
\begin{equation}\label{guesscrosscap}
|C:[\mu],S\rangle \equiv {\cal P}_o
|C:\mu,S\rangle = \frac{(1+g+ \cdots + g^K)}{\sqrt{K}}\
|C:\mu,S\rangle =\frac1{\sqrt{K}} \sum_{\nu_0}^{K-1} 
|C:\mu+\nu_0\mu_0,S+2\nu_0\rangle\
.
\end{equation}
where $\mu_0\equiv(\stackrel{r\ \text{times}}{1,\cdots,1})$. 
However, as we have seen, the $\BZ_{K}$ action is projectively realised
on crosscap states (see eqn. (\ref{symmcs})). This implies that one needs
to introduce suitable phases into the above expression. 
The orbifolded crosscap state is given by\footnote{The general analysis
of crosscap states has been carried in ref. \cite{BH1} and seems to suggest
the addition of additional phases in the crosscap states. However, we have
taken the practical view of looking to directly implement in the $\beta_0$
projection, i.e., keep only Ishibashi states that have  odd integral  $U(1)$
charge.}
\begin{equation}\label{fullcrosscap}
\boxed{
|C:[\mu],S\rangle = \sum_{\nu_0=0}^{K-1} \frac{e^{i\pi\nu_0}}{\sqrt{K}} 
\Big[\frac{\overline{\sigma}_{\mu+2\nu_0\mu_0;S+2\nu_0}}{
\overline{\sigma}_{\mu,S}}\Big]\ 
|C:\mu+2\nu_0\mu_0,S+2\nu_0\rangle
}
\end{equation}
where 
$\overline{\sigma}_{\mu,S}\equiv \prod_i \sigma_{0,M_i,S}$.
Since all the crosscaps from the
individual models have the same S, the spacetime part is uniquely fixed.
So the index $(S+2\nu_0)$, indicates that the spacetime part is shifted as
well.
One can put in the explicit form of the crosscap state in the individual
minimal model and see that Ishibashi states whose total $U(1)$ charge is
not odd integral are projected out. So the $\beta_0$ projection has been
carried out.

One can check that
the following properties are true (atleast for odd $k_i$)
\begin{equation}\label{shiftcrosscap}
\begin{split}
|C:[\mu+2\mu_0],S+2\rangle &= -
\Big[\frac{ \overline{\sigma}_{\mu,S}}{\overline{\sigma}_{\mu+2\mu_0;S+2}}
\Big]\
|C:[\mu],S\rangle \\
|C:[\mu],S+4\rangle &= (-)^{S r}\ |C:[\mu],S\rangle
\end{split}
\end{equation}

For even $S$, the crosscap states consists of NSNS Ishibashi states alone
and for odd $S$, they consist of RR Ishibashi states alone. Clearly,
this cannot provide a supersymmetric crosscap. The basic reason is that 
using either do not produce the produce the full Klein-bottle partition 
function of a supersymmetric theory -- using NSNS part only reproduce the
characters without $\wi$-twisting, \viz $NS$ and $R$, while using RR
part only reproduce the characters with $\wi$-twisting only, \viz 
$\wt{NS}$ and $\wt{R}$. Hence, the supersymmetric
crosscap is a linear combination of the two possibilities. The spectral
flow generator is the product of the spectral flow generators, $f_i$,
in the individual models and let $f_0$ indicate the same in the spacetime
sector. The full crosscap in the Gepner model is given by the orbifolding
by the group $f$. However, since $f^2=gh$, this has a $\BZ_2$ action on the
crosscap states already constructed. We obtain
\begin{equation}\label{gepnercrosscap}
\boxed{
|C:[\mu]\rangle_{\text{Gepner}} \equiv  \frac1{\sqrt2}\Big(|C:[\mu],0\rangle
\pm |C:[\mu+\mu_0],1\rangle\Big)
}
\end{equation}
This is our proposal for the crosscap state for Gepner models. 
The two signs reflect the possibility two having orientifold
planes and anti-orientifold planes, since the signs of RR charges are
opposite.

\sectiono{A Toy Model : $(k\,=\,1)^3$ Gepner Model}\label{ns1}
\medskip

As a warm-up to orientifold of Calabi-Yau 3-fold we start with the Gepner 
model realization of the simplest Calabi-Yau, \viz {\rm CY}$_1$. This is 
nothing but compactification on $T^2$ at some special values of its 
moduli. Let the complex structure and K\"ahler moduli of $T^2$ be denoted 
by $\tau$ and $\rho$. Since compactification on $T^2$ corresponds to 
$c\,=\,3$, it can be realized as 3 types of Gepner models\cite{TDUALITY}, 
\viz 
$(k\,=\,1)^3$, $(k\,=\,2)^2$ and $(k\,=\,1)\cdot(k\,=\,4)$. All these 
Gepner models correspond to $T^2$ compactification at the enhanced 
symmetry points, \eg, the former corresponds to $T^2$ compactification at 
$SU(3)$ point with ($\tau$,$\rho$) = ($e^{2\pi i/3}$, $e^{2\pi i/3}$), the 
second on $T^2$ at the $SU(2)^2$ point with ($\tau$, $\rho$) = ($i$, $i$) 
and the latter on $T^2$ at $SU(3)$ point with ($\tau$, $\rho$) = ($e^{2\pi 
i/3}$, $e^{2\pi i/3}$). For simplicity we consider the $(k\,=\,1)^3$ 
Gepner model which is equivalent to compactification on an $SU(3)$ torus 
with $(\tau\,,\rho)\,=\,(e^{2\pi i/3}\,,\,e^{2\pi i/3})$; the $SU(3)$ 
torus is generated by quotienting $\RR^2$ by $SU(3)$ root lattice. 
We now discuss the Gepner model and construction of crosscap states 
therein.
\medskip 

\subsection{A-Type Crosscap States of $(k\,=\,1)^3$ Model}
\label{nss1.1}
\medskip

In this section we discuss the orientifold of $(k=1)^3$ Gepner model. We 
consider type IIA theory compactified on this Gepner model. We discuss the 
symmetry group of this model and construct the orientifold model from RCFT
approach. Our main goal is to write all possible crosscap states in this 
orientifold model and compute their overlaps and hence the KB amplitudes. 
Using the abstract formalism of section \ref{bs4}, we have already written 
down the formula for crosscaps in Gepner model in section \ref{cs1}. At 
the final step, we need to know the precise orientifold group for the model.
This allows us to consider crosscaps arising from the simple currents. These
simple currents are due to the presence of discrete orbifold group present
in the full orientifold group. We discuss this issue in the next section.

\subsubsection{The Crosscaps}\label{nsss1.3}
\medskip

Once we know the orientifold group of the theory, it is quite straightforward
to write down the crosscap states. Since for $(k=1)^3$ model we know the 
orientifold group is as given by eqn.\refb{og}, we immediately infer 
that
this model has 9 crosscap states. In the large volume limit, when we have the
geometric phase of Calabi-Yau, these crosscaps correspond to
 $\RP^1$ cycles.

We first write down the NSNS part of the crosscaps in the theory.
As prescribed in eqn.\refb{b2.29}, we set $L_1=L_2=L_3=0$. To determine 
$[\mu] = (M_1,M_2,M_3)$ in eqn.\refb{gepnercrosscap}, 
we apply the machinery of section \ref{bss2.2}. The simple 
currents which generate these 9 crosscaps are obtained 
by tensoring the currents $\Phi_{000}$, $(\Phi_{011})^2=\Phi_{022}$ and 
$(\Phi_{011})^4=\Phi_{044}$ of the $(k=1)^3$ theory\fn{It can be obtained
by another way. As we see from eqn.\refb{gepnercrosscap} and 
section \ref{bss2.2}, we set $L_i=S_i = 0$ for these states. The only relevant
labels for these states are $M_i$ which take values $0,2,4 \mod 6$ \ie 27
values. It becomes 9 after imposing the equivalence $\mu
\leftrightarrow \mu + 2+\mu_0$.}.
Thus NSNS part of the crosscap state of this model can be denoted as
\begin{equation}\label{n12}
\boxed{
\Ket{C\,: M_r = 2a_r}_{NSNS}\,\equiv\,
\Ket{C\,: [\mu] = [2a]}_{NSNS}\,,\quad 
a_r\,=\,0,1,2
}
\end{equation}

After obtaining the NSNS part, it is now easy to get the RR part. The general
prescription is to spectral flow the NSNS boundary state by $f$, as we 
discussed in section \ref{css2}. So following eqn.\refb{gepnercrosscap}, the
total crosscap state in the Gepner model $(k=1)^3$ is
\begin{equation}\label{n12a}
\Ket{C\ :\  [\mu]\ ;\  (k=1)^3}_{\text{Gepner}} = \frac{1}{\sqrt 2}
\Bigg[\;\Ket{C\ :\  [\mu] = [2a]}_{\text{NSNS}} + 
\Ket{C\ :\  [\mu + \mu_0], 1}_{\text{RR}}\Bigg]
\end{equation}

\subsection{Spectra, Character Formulae and Spectral Flow Invariant Orbits 
of $k\,=\,1$ Minimal Model}\label{nss1.2}

Before we go and discuss the KB amplitude in this model, we need to know the
spectra of the $k\,=\,1$ minimal model and how to express it neatly in terms
of spectral flow invariant orbits of supersymmetric characters -- a technique
introduced in ref.\cite{EOTY}. This is a very powerful technique for 
discussing Gepner model and in fact, we shall express the KB amplitude in
this model in terms of these orbits.

The NS-sector of $k\,=\,1$ minimal model contains three representations as 
given in table 3 along with their conformal weights and $U(1)$ 
charge.
\renewcommand{\arraystretch}{1.8}
\begin{table}\label{F.1}
\bec
\begin{tabular}[!hb]{||c|c|c|c|c||}\hline
 & $l$ & $m$ & $h$ & $q$\\[-3pt]\hline
$A_+$ & 0 & 0 & 0 & 0 \\[-3pt]\hline
$B_+$ & 1 & 1 & $\frac{1}{6}$ & $\frac{1}{3}$ \\[-3pt]\hline
$C_+$ & 1 & $-$1 & $\frac{1}{6}$ & $-\,\frac{1}{3}$\\ \hline
\end{tabular}
\eec
\caption{NS representations of the $k\,=\,1$ minimal model and their 
characters.} 
\end{table}

Under $\eta\,\to\,\eta + 1$, the spectral flow amongst the three 
types of characters are given by
\begin{equation}\label{F.2} 
A_{\pm}\;\xto{}\;B_{\pm}\;\xto{}\;C_{\pm}\;\xto{}\;A_{\pm}
\end{equation}
The formulae of the NS- and R-sector characters in terms of the string 
functions and theta functions can be found in eqns.\refb{F.4}, \refb{F.5}
and \refb{F.6} in the appendix \ref{A6}.

The spectral flow invariant orbits for $(k\,=\,1)^3$ 
are really very trivial\cite{EOTY,GJS}. 
This model has the following two spectral flow invariant 
orbits in NS sectors :
\begin{enumerate}
\item ``Massless Orbit'' :
\begin{equation}\label{F.8}
\NS^+_0\,=\,A^3_+\,+\,B^3_+\,+\,C^3_+
\end{equation}
We call this orbit the ``massless'' orbit, since it gives rise to 
 graviton($h\,=\,q\,=\,0$) and massless matters($h\,=\,\half$, $q\,=\,\pm 1$) 
in spacetime.
\item ``Massive Orbit'' :
\begin{equation}\label{F.9}
\NS^+_1\,=\,3\,A_+\,B_+\,C_+
\end{equation}
This orbit is called the ``massive'' orbit, since it gives rise to massive 
matter in spacetime.
\item There are two other orbits corresponding to $\NS_0^+$ and $\NS_1^+$
which appear in the $\wi$ twisted sectors :
\begin{equation}\label{F.9a}
\NS_0^- = A^3_- + B^3_- + C^3_-\,\quad,\quad \NS_1^- = 3\,A_-B_-C_-
\end{equation}
\end{enumerate}

\medskip

\subsection{KB Amplitude}\label{nss1.3}
\bigskip

We now compute the KB amplitude in our model using the ansatz given in 
eqns.\refb{fullcrosscap} and \refb{gepnercrosscap}. It is given by sum 
of two pieces : an NSNS amplitude and 
a RR amplitude;
\ben\label{n16}
& & 
\Bra{C\,:\,[\wt{\mu}]}
e^{\frac{-\pi i}{2\tau}H_{cl}}\,
\Ket{ C\,:\,[\mu]}\non\\ 
&=&
{\Big. \Big.}_{\text{NSNS}}
\Bra{C\,:\,[\wt{\mu}] = [2\wt{a}],0}
e^{\frac{-\pi i}{2\tau}H_{cl}}\,
\Ket{ C\,:\,[\mu] = [2a],0}_{\text{NSNS}}\non\\
&+& {\Big. \Big.}_{\text{RR}}
\Bra{C\,;\, [2\wt{a} + \mu_0], 1}
e^{\frac{-\pi i}{2\tau}H_{cl}}\,
\Ket{ C\,:\,[2a + \mu_0]}_{\text{RR}}
\een

\bigskip

\noi$\bullet\;\;${\bf NSNS part of the KB Amplitude :}
\bigskip

\noindent 
Using the identity $Y^0_{l_i\ 0} = (-1)^{l_i}$, the NSNS amplitude is given by
\ben\label{nskb}
\KB_{\text{NSNS}} &=& {\Big. \Big.}_{\text{NSNS}} 
\Bra{C\,:\,[2\wt{a}],0}
e^{-\frac{\pi i H}{2\tau}}\,\Ket{ C\,:\, [2a],0}_A \nonumber\\
&=&
\frac1K\sum_{\tilde{\nu}_0=0}^{K-1} \sum_{\nu_0=0}^{K-1}\
\delta^{2}_{[\mu]+[\tilde{\mu}]}\
e^{-i\pi \nu_0}\ e^{+i\pi \tilde{\nu}_0}\ 
\ol{\sigma}_{0,2\wt{a},0}\ \ol{\sigma}_{0,2a,0} \non\\
&\times& \half\ \big(\chi^{-\nu_0}_{\text{st}}\ +\ 
\chi^{-\nu_0+ 2}_{\text{st}}\big)\ \prod_i 
\sideset{}{^{\text{ev}}}\sum_{l_i} (-)^{l_i} 
\Big[
\chi_{l_i,\frac{\tilde{M_i}-M_i}2-(\nu_0 - \tilde{\nu}_0)}^{
(- (\nu_0 - \tilde{\nu}_0))}(2\tau)
+ \chi_{l_i,
\frac{\tilde{M_i}-M_i}2- (\nu_0 - \tilde{\nu}_0)}^{(-(\nu_0 - 
\tilde{\nu}_0)+2)}(2\tau)\Big]\,,\non\\
& &
\een
where 
$\chi_{\text{st}}$ is the appropriate spacetime character. In 
evaluating 
the sum over $\nu_0$ and $\tilde{\nu}_0$, it is better to
to shift $\nu_0$ by $\tilde{\nu}_0$, this shift does not make any
difference since the arguments are periodic.

\noindent
The final form of NSNS part of the KB amplitude is
\ben\label{nKB}
\KB_{\text{NSNS}} &=& {\Big. \Big.}_{\text{NSNS}} 
\Bra{C\,:\,[2\wt{a}],0}
e^{-\frac{\pi i H}{2\tau}}\,\Ket{ C\,:\, [2a],0}_{\text{NSNS}} \nonumber\\
&=&
\sum_{\nu_0=0}^{K-1}\ 
e^{-i\pi\nu_0}
\delta^{2}_{[\mu]+[\tilde{\mu}]}\
\ol{\sigma}_{0,2\wt{a},0}
\ol{\sigma}_{0,2a,0}\non\\
&\times& \half\ \big(\chi^{-\nu_0}_{\text{st}}\ +\ 
\chi^{-\nu_0+2}_{\text{st}}\big)\ \prod_i 
\sideset{}{^{\text{ev}}}\sum_{l_i} (-)^{l_i} 
\Big[
\chi_{l_i,\frac{\tilde{M_i}-M_i}2-\nu_0}^{
(-\nu_0) }
+ \chi_{l_i,
\frac{\tilde{M_i}-M_i}2-\nu_0}^{(-\nu_0+2)}\Big]\big(2\tau\big)\,.
\een
Eqn.\refb{nKB} is the {\em master formula} for KB amplitude in NSNS 
sector. The above formula holds in general (as long as all the $k_i$ are
odd) and thus holds for the $(k\,=\,1)^3$ and $(k\,=\,3)^5$  Gepner
models.
Thus, we have kept eqn.\refb{nKB} completely generic. Here $K\,=\,{\rm 
lcm}\,(4\,,\,2\,k\,+\,4)$ and $p$ is the number of minimal models 
being tensored. For $k\,=\,1$, $K\,=\,12$ and $p\,=\,3$. 


Using the formulae \refb{F.8}, \refb{F.9} for the flow 
invariant orbits, we can finally express the KB amplitudes for crosscaps
for $(k\,=\,1)^3$ Gepner model 
in terms these orbits.  The answer for the case of amplitudes involving the
same crosscap state is
\ben\label{n18}
& &
{\Big. \Big.}_{\text{NSNS}}
\Bra{C\,:\,[\mu] = [2a]}e^{\frac{-\pi i}{2\tau}H_{cl}}\,
\Ket{C\,:\, [\mu] = [2a]}_{\text{NSNS}}\non\\
&=& \Bigg[\chi^{+,\text{(NS)}}_{\text{st}}\;
\Big(\;\prod_{r\,=\,1}^3\;
\big(\chi^0_{00}\,+\,\chi^2_{00}\big)(2\tau)\,+\;\prod_{r\,=\,1}^3\;
\big(\chi^0_{11}\,+\,\chi^2_{11}\big)(2\tau)\non\\ 
&+& \prod_{r\,=\,1}^3\;
\big(\chi^{0}_{1-1}\,+\,\chi^2_{1-1}\big)(2\tau)\;\Big)\
-\ \chi^{+,\text{(R)}}_{\text{st}}\;\Big(\prod_{r\,=\,1}^3\;
\big(\chi^1_{01}\,+\,\chi^{-1}_{01}\big)(2\tau)\,+\;\prod_{r\,=\,1}^3\;
\big(\chi^1_{10}\,+\,\chi^{-1}_{10}\big)(2\tau)\non\\ 
&+& \prod_{r\,=\,1}^3\;
\big(\chi^{1}_{12}\,+\,\chi^{-1}_{12}\big)(2\tau)\Big)\Bigg]\non\\
&=&\Bigg[\chi^{+,\text{(NS)}}_{\text{st}}\;
\Big(A^3_+\,+\,B^3_+\,+\,C^3_+\Big)\ -\ 
\chi^{+,\text{(R)}}_{\text{st}}\;
\Big(\wh{A}^3_+\,+\,\wh{B}^3_+\,+\,\wh{C}^3_+\Big)\Bigg]\,(2\tau)\,,
\een
Thus the NS-part of the KB amplitudes in
$(k\,=\,1)^3$ Gepner model is schematically(\ie suppressing the Minimal 
model labels) 
\begin{eqnarray}\label{n20}
\text{KB}^{(k=1)^3}_{\text{NS}} &=& 
\chi^{+,\text{(NS)}}_{\text{st}}\;
(A^3_+\,+\,B^3_+\,+\,C^3_+)(2\tau) \ -\ 
\chi^{+,\text{(R)}}_{\text{st}}\;
(\wh{A}^3_+\,+\,\wh{B}^3_+\,+\,\wh{C}^3_+)(2\tau)\non\\ 
&=& 
\chi^{+,\text{(NS)}}_{\text{st}}\ \NS_0^+(2\tau) \ -\ 
\chi^{+,\text{(R)}}_{\text{st}}\ \RA_0^+(2\tau)
\end{eqnarray}
This eqn.\refb{n20} is the desirable result, as it agrees with the 
general structure of the NSNS part of the KB amplitude. As discussed 
after eqn.\refb{kbmm}, the NSNS part of crosscap state 
reproduces the characters $NS$ and $R$ in the loop channel, \ie
\ben\label{n20a}
& & {\Big. \Big.}_{\text{NSNS}}\Bra{C\,; [\mu]}
e^{\frac{-\pi i}{2\tau} H_{cl}}
\Ket{C\ :\  [\mu]}_{\text{NSNS}}\non\\
&=& \half\,\Bigg[{\rm Tr}_{{\rm NSNS}}\Big[\,\Omega\cdot\sigma\cdot\SP\,
e^{2\pi i\tau\,H_{cl}}\Big]\,-\,
{\rm Tr}_{{\rm RR}}\Big[\,\Omega\cdot\sigma\cdot\SP\cdot
e^{2\pi i\tau\,H_{cl}}\Big]\Bigg]\non\\
&\equiv& NS\ -\ R
\een
Thus from eqns.\refb{n20} and \refb{n20a}, we find that for overlaps of
crosscaps,
\begin{equation}\label{n20b}
NS = \chi^{+,\text{(NS)}}_{\text{st}}\ \NS_0^+(2\tau)\,,\qquad 
R = \chi^{+,\text{(R)}}_{\text{st}}\ \RA_0^+(2\tau)
\end{equation}
Here $\SP$ denote the projector for
the subgroup $\HH=\BZ^3_3/\BZ_3$.
The sum $\wi$ twisted characters, 
actually comes from the overlap of the RR-part of the 
crosscap state\cite{POLCHINSKICAI}.

\noi{\bf RR part of the KB Amplitude :}
\bigskip

We can repeat the same analysis for the RR sector amplitude. We find
\ben\label{n20c}
\text{KB}_{\text{RR}} &=& {\Big. \Big.}_{\text{RR}}
\Bra{C\,:\ [2\wt{a} + \mu_0], 1}
e^{\frac{-\pi i}{2\tau} H_{cl}}
\Ket{C\ :\  [2a + \mu_0], 1}_{\text{RR}}\non\\
&=& \Bigg[\chi^{-,\text{(NS)}}_{\text{st}}\;
\Big(A^3_- + B^3_- + C^3_-\Big)\ - \
\chi^{-,\text{(R)}}_{\text{st}}\;
\Big(\wh{A}^3_-\,+\,\wh{B}^3_-\,+\,\wh{C}^3_-\Big)\Bigg]\ (2\tau)\non\\
&=& \chi^{-,\text{(NS)}}_{\text{st}}\;\NS_0^-(2\tau) \ -\  
\chi^{-,\text{(R)}}_{\text{st}}\;\RA_0^-(2\tau)
\een
Since we expect
\ben\label{n21}
& & {\Big. \Big.}_{\text{RR}}
\Bra{C\,:\ [2\wt{a} + \mu_0], 1}
e^{\frac{-\pi i}{2\tau} H_{cl}}
\Ket{C\ :\  [2a + \mu_0], 1}_{\text{RR}}\non\\
&=& \half\,\Bigg[{\rm Tr}_{{\rm NSNS}}\Big[\,\Omega\cdot\sigma\cdot\SP\,\wi\,
e^{2\pi i\tau\,H_{cl}}\Big]\,-\,
{\rm Tr}_{{\rm RR}}\Big[\,\Omega\cdot\sigma\cdot\SP\cdot\wi\cdot
e^{2\pi i\tau\,H_{cl}}\Big]\Bigg]\non\\
&\equiv& \wt{NS} \ -\ \wt{R}\,,
\een
in the loop channel, we find from eqns.\refb{n20b} and \refb{n21} that for 
RR overlaps 
between crosscaps
of this Gepner model 
\begin{equation}\label{n22}
\wt{NS} = \chi^{-,\text{(NS)}}_{\text{st}}\ \NS_0^-(2\tau)\,,\qquad 
\wt{R} = \chi^{-,\text{(R)}}_{\text{st}} \RA_0^-(2\tau)
\end{equation} 

\noindent
Summing eqns.\refb{n20} and \refb{n20c} we get the total KB amplitude in
the orientifold of $(k=1)^3$ Gepner model :
\begin{eqnarray}\label{KBTOTAL}
\text{KB}^{(k=1)^3}_{\text{total}} &=&
(NS + \wt{NS}) - (R + \wt{R})\non\\
&=& \Big(\chi^{+,\text{(NS)}}_{\text{st}}\ \NS^+_0 \ +\ 
\chi^{-,\text{(NS)}}_{\text{st}}\ \NS^-_0\Big)(2\tau) \ - \
\Big(\chi^{+,\text{(R)}}_{\text{st}}\ \RA^+_0 \ +\ 
\chi^{-,\text{(R)}}_{\text{st}}\ \RA^-_0\Big)(2\tau)\non\\
& & 
\end{eqnarray}

The result involving different crosscaps which however preserve the same
supersymmetry is non-zero and involves the 
{\em massive} orbits.
In this example, it is $3ABC$ and characters obtained from this by
means of spectral flow.

\sectiono{KB Amplitudes in $(k\,=\,3)^5$ Gepner Model}
\label{qs1}
\medskip

Finally we come to the point of computing the KB amplitudes in the 
orientifolds of the quintic. As discussed in section \ref{cs1}, 
compactification on
the quintic hypersurface can be represented by tensoring five copies of
$k\,=\,3$ minimal model. This is Gepner's famous $(k\,=\,3)^5$ model
\cite{GEPNER}. The formula for the KB amplitudes can be written straight away
using the master formula eqn.\refb{n20} derived in section \ref{nss1.3} by
substituting the relevant values for $p$ and $K$. Next problem is to 
manipulate this expression so as to express it in terms of orbits of 
$k\,=\,3$ minimal model and we can derive something meaningful from it. Our
previous experience with $(k\,=\,1)^3$ model shows that the answer should be
proportional to the massless orbits.This is also expected from hindsight, 
since only the massless tadpoles flow in the loop channel.

So we need to know the representations of $k\,=\,3$ minimal model. This can
be worked out along the lines of $k\,=\,1$ model. The only non-trivial part
is to work out the spectral flow invariant orbits which we will discuss here.
To work out these orbits, it is quite useful to see what $(2\,,2)$ SCFT 
teaches us about this spectra and the massless and massive representation.
Note that before taking the orientifold projection the worldsheet algebra was 
given by $(2\,,\,2)$ SCFT.

\bigskip

\subsection{(2,2) SCFT Spectrum}\label{qss1.2}
\medskip

To work out the spectrum for compactification with (2,2) SCFT, the 
knowledge of $\NN\,=\,2$ SCA is not enough. In fact, we already noticed 
that this SCA has an automorphism generated by the spectral flow operator
$U_{\,\etal\,,\etar}$, eqn.\refb{SCAspectralflow}. It is well-known that 
under the spectral flow, NS- and R-sectors are transformed into each 
other, so that the spacetime bosons and fermions are paired. To work out 
the massless and massive spectrum of (2,2) SCFT with $c\,=\,9$, one has to 
incorporate the spectral flow operator into the $\NN\,=\,2$ 
SCA\cite{EOTY}. This 
makes the latter into a bigger algebra; it is not of the kind of Lie 
algebra but a kind of $W$-algebra. This algebra can be worked out as 
follows. Representations of $\NN\,=\,2$ SCA are labeled by the conformal 
weight $h$ and the $U(1)$ charge $q$. In the case of $c\,=\,3\,n$, where 
$c$ is the central charge of the Virasoro algebra, the vacuum 
state($h\,=\,0$, $q\,=\,0$) in the NS sector is mapped onto the states
($h\,=\,\dfrac{n}{8}$, $q\,=\,\pm \dfrac{n}{2}$) in the R-sector and
($h\,=\,\dfrac{n}{2}$, $q\,=\,\pm n$) in the NS-sector under the spectral 
flow $\eta\,=\,\pm \half\;,\pm 1$ respectively. The states with 
($h\,=\,\dfrac{n}{8}$, $q\,=\,\pm \dfrac{n}{2}$) and 
($h\,=\,\dfrac{n}{2}$, $q\,=\,\pm n$) are respectively the covariantly 
constant spinors and (anti-)holomorphic $n$-forms of $\CY_n$ with $SU(n)$ 
holonomy. The former corresponds to the spacetime supersymmetry and the 
latter to the flow generators\cite{EOTY}. We denote the operators 
corresponding to flow generators by $U(z)$ and $\ol{U}(\zbar)$ and its 
superpartners with ($h$, $q$) = ($\dfrac{n + 1}{2}$, $\pm (n - 1)$) by 
$V(z)$ and $\ol{V}(\zbar)$. For compactification on $CY_3$, 
we have $c\,=\,9$ hence $n\,=\,3$ and the flow generators are states 
labeled by ($h\,=\,\threehalf$, $q\,=\,\pm 3$).

This bigger algebra had been worked out for $c\,=\,9$ by 
Odake\cite{ODAKE}. We call it {\em Odake Algebra} 
and list the values of $h$ and $q_L$ for left-moving massless 
and massive spectra in table 4. Similarly for right-moving 
sectors.

\medskip
\renewcommand{\arraystretch}{1.3}
\begin{table}[!ht]\label{SCA2}
\bec
\begin{tabular}{||c|c|c||}\hline
 & NS & R \\\hline
 & $\wh{\text{NS}}_1$ : $h$ = 0, $q_L$ = 0 & $\wh{\text{R}}_1$ : $h$ = 
$\frac{3}{8}$, 
$q_L$ = $\pm\,\frac32\;\;$ (Identity operator)\\\hline
Massless & $\wh{\text{NS}}_2$ : $h$ = $\frac12$, $q_L$ = 1 & 
$\wh{\text{R}}_2$ : 
$h$ = ${\frac38}$, 
$q_L$ = $-\,{\frac12}\;\;$ ($\etal\,=\, -\,{\frac12}$)\\
 &  $\wh{\text{NS}}_3$ : $h$ = $\frac12$, $q_L$ = $-$1 & 
$\wh{\text{R}}_3$ :$h$ = 
$\frac38$, 
$q_L$ = ${\frac12}\;\;$ ($\etal\,=\,{\frac12}$)\\\hline
Massive & $\wh{\text{NS}}_4$ : $h\,>\,0$, $q_L$ = 0 & $\wh{\text{R}}_4$ : 
$h\,>\, 
{\frac38}$, 
$q_L$ = $\pm\,{\frac32}$, $\pm\,{\frac12}$\\
& $\wh{\text{NS}}_5$ : $h\,>\, {\frac12}$, $q_L\,=\, \pm\,1$ & 
$\wh{\text{R}}_5$ : 
$h\,>\, {\frac38}$,$q_L\,=\,\pm\,{\frac12}$\\\hline
\end{tabular}
\eec
\caption{Dimensions and $U(1)$ charges of the massless and 
massive spectra of the holomorphic part of the (2,2) internal SCFT. 
Similar results holds for the antiholomorphic part.}
\end{table}

It shows that there are mainly 3 massless and 2 massive orbits in $c\,=\,9$ 
SCFT. It helps us immensely in constructing the orbits for $(k\,=\,3)^5$ 
Gepner model. We must make a note of one point. The Odake algebra gives rise
to an {\it irrational} CFT, since it contains {\it infinite} number of
primaries. But the same $c\,=\,9$ SCFT can be constructed as an RCFT, as \eg, 
in terms of $(k\,=\,3)^5$ Gepner model. So one might wonder about the 
connection between the Gepner model and the abstract $c\,=\,9$ Odake algbera
and its representation. In fact, if we perform an infinite sum over the 
characters of the latter, that organise itself nicely into the orbits of
Gepner model. This point was emphasized in ref.\cite{EOTY,ODAKE} and they
explicitly checked it for a few other Gepner models.

\bigskip

\subsection{Spectra and Spectral Flow Invariant Orbits of $k\,=\,3$ 
Minimal Model}\label{qss1.3}

The NS-sector of $k\,=\,3$ minimal model consists of ten representations 
as given in table \ref{G.1} of appendix \ref{A7}. We also provide their 
conformal weights and $U(1)$ charges.
From the table \ref{G.1}, we find that ten representations of $k\,=\,3$ 
minimal model forms two non-overlapping groups of five each under the 
spectral flow $\eta\,\to\,\eta\,+\,1$. They are 
\begin{equation}\label{G.2}
A_+\,\xto{}\,J_+\,\xto{}\,I_+\,\xto{}\,H_+\,\xto{}\,G_+\,\xto\,A_+\,,
\end{equation}
\begin{equation}\label{G.3}
B_+\,\xto{}\,F_+\,\xto{}\,E_+\,\xto{}\,D_+\,\xto{}\,C_+\,\xto{}\,B_+
\end{equation}

The expressions of $k\,=\,3$ NS-sector characters 
can be found in eqns.\refb{G.1a} in appendix \ref{A7}.

We now write down the most obvious four spectral flow invariant orbits in 
the NS sector of the model. 
\begin{enumerate}
\item ``Graviton Orbit''
\begin{equation}\label{q2}
\NS^+_0\,=\,A_+^5\,+\,G_+^5\,+\,H_+^5\,+\,I_+^5\,+\,J_+^5
\end{equation}
It contains the operator for the spacetime graviton($h\,=\,q\,=\,0$).
\item ``Self-conjugate Massless Matter Orbit''
\begin{equation}\label{q3}
\NS^+_1\,=\,B_+^5\,+\,C_+^5\,+\,D_+^5\,+\,E_+^5\,+\,F_+^5
\end{equation}
It contains the operators for both hyper and vector multiplets($h\,=\,0$,
$q\,=\,\pm 1$) and hence the name.
\item ``Massive Orbits''
\begin{equation}\label{q4}
\NS^+_2\,=\,5\,A_+\,G_+\,H_+\,I_+\,J_+\;,\quad
\NS^+_3\,=\,5\,B_+\,C_+\,D_+\,E_+\,F_+
\end{equation}
Both $\text{NS}_2$ and $\text{NS}_3$ are ``massive'' orbits.
\item
There are $\wi$ twisted orbits corresponding to $\NS_0^+$, $\NS_1^+$, 
$\NS_2^+$ and $\NS_3^+$. They are
\ben\label{q2a}
\NS^-_0 &=& A_-^5\,+\,G_-^5\,+\,H_-^5\,+\,I_-^5\,+\,J_-^5\non\\
\NS^-_1 &=& B_-^5\,+\,C_-^5\,+\,D_-^5\,+\,E_-^5\,+\,F_-^5\non\\
\NS^-_2 &=& 5\,A_-\,G_-\,H_-\,I_-\,J_-\,,\quad
\NS^-_3 \,=\, 5\,B_-\,C_-\,D_-\,E_-\,F_-
\een
\end{enumerate}
These are not the set of all orbits for this model. There are other 
massive orbits. Since we do not require them, we are not listing them here. 

The orbits in R-sector of this model, in particular, $\RA_0^{\pm}$, 
$\RA_1^{\pm}$, $\RA^{\pm}_2$, $\RA^{\pm}_3$ and $\RA^{\pm}_4$ are 
obtained from eqns. \refb{q2}, \refb{q3}, \refb{q4} and \refb{q2a} by 
replacing  the supersymmetric characters by their hatted counterpart 
as defined in eqn. \refb{G.1c} in appendix \ref{A7}.

\bigskip

\subsection{KB Amplitudes}\label{qss1.4}
\medskip

In this section we are going to compute the KB amplitudes for $(k=3)^5$ 
Gepner model corresponding to quintic. 
Since the
orientifold group for this model is as given in eqn.\refb{og}.
\fn{Compared to $(k=1)^3$ model, for this 
model we need the the factor of $\wil$ in the orientifold group.},
clearly it gives rise to $5^4\,=\,625$ crosscaps states which in the large
volume limit corresponds to 625 $\RP^3$'s. In the Gepner model such crosscaps
are constructed as follows. All such 625 crosscaps are actually generated by
the symmetry group elements of $\HH$ from the basic one \ie the crosscap 
with $L_i=M_i=S_i=0$. So we can apply the techniques of section 
\ref{bss2.2},
in particular, we apply the formulae \refb{b2.35} and \refb{b2.36}. It is
easy to figure out that these 625 crosscaps are generated by tensoring the
simple currents $\Phi_{000}$, $(\Phi_{011})^2=\Phi_{022}$,
$(\Phi_{011})^4=\Phi_{044}$, $(\Phi_{011})^6=\Phi_{066}$ and
$(\Phi_{011})^8=\Phi_{088}$. It implies that we should put $M_i=0,2,4,6,8
\mod 10$
in eqn.\refb{c1} or eqn.\refb{c2} in each crosscap state of five copies 
of minimal model.
Thus the NSNS part of a  generic crosscap state(A-type) in this model would be
\begin{equation}\label{q5c}
\Ket{C\,;\, [\mu] = [2a]}_{\text{NSNS}}\,,\quad a=0,2,4,6,8\mod 10
\end{equation}
where $a_r=0,2,4,6,8$. The RR part of the crosscap state is 
\begin{equation}\label{q5rr}
\Ket{C\,;\, [\mu] = [2a],1}_{\text{RR}}\,,\quad a=0,2,4,6,8\mod 10\,.
\end{equation} 
The full
crosscap state of the Gepner model is the sum of the NSNS part 
and the RR part with Gepner projection applied as in 
eqns.\refb{fullcrosscap} and \refb{gepnercrosscap}.

All remains is to compute the KB amplitude using the formulae for NSNS and RR
part of the crosscaps. The master formula had already been worked out in eqn.
\refb{nKB}. Since we have found the flow invariant orbits of the 
model, it is easy to write down the formula for
the KB amplitude in  the orientifold of $(k\,=\,3)^5$ Gepner model. Putting
$p\,=\,5$ and $K\,=\,20$ in eqn.\refb{nKB} and explicitly evaluating it, 
we find that for $\mu=\tilde{\mu}$
we get\fn{Here we define $NS = \half\  
\text{Tr}_{\text{NSNS}}[\Omega\cdot\wil\cdot\sigma\cdot{\sf P}\cdot 
e^{2\pi i\tau H}]$ and $\wt{NS} = \half\ 
\text{Tr}_{\text{NSNS}}[\Omega\cdot\wil\cdot\sigma\cdot{\sf 
P}\cdot\wi\cdot
e^{2\pi i\tau H}]$; similarly the R-sector characters.} 
\ben\label{q6}
\text{KB}^{(k=3)^5}_{\text{NS}} &=& NS - R\non\\
&=& \Bigg[\;
\half\ \chi^{+, \text{(NS)}}_{\text{st}}\Big\{(A^5_+ + G^5_+ + H^5_+
+ I^5_+ + J^5_+)
\ +\ (B^5_+ + C^5_+ + D^5_+ + E^5_+ + F^5_+)\Big\}(2\tau)\non\\
&-& \half\ \chi^{+, \text{(R)}}_{\text{st}}\Big\{
(\wh{A}^5_+ + \wh{G}^5_+ + \wh{H}^5_+
+ \wh{I}^5_+ + \wh{J}^5_+)
\ +\ (\wh{B}^5_+ + \wh{C}^5_+ + \wh{D}^5_+ + \wh{E}^5_+ + \wh{F}^5_+)\Big\}
(2\tau)\ \Bigg]
\non\\
&=& \chi^{+,\text{(NS)}}_{\text{st}}\ (\NS^+_0\,+\,\NS^+_1) \ -\  
\chi^{+,\text{(R)}}_{\text{st}}\ (\RA_0^+ + \RA_1^+)
\een
This is the desired result, since it shows that only the orbits of massless 
tadpoles circulate in the loop channel. Similarly the RR amplitudes for
crosscaps satisfying $[\mu] = [\wt{\mu}]$ is

\ben\label{q7}
\shoveright{\text{KB}^{(k=3)^5}_{\text{RR}}} &=& \wt{NS} - \wt{R}\non\\
&=& \Bigg[
\half\ \chi^{-,\text{(NS)}}_{\text{st}}\ \Big\{(A^5_- + G^5_- + H^5_-
+ I^5_- + J^5_-)
\ +\ (B^5_- + C^5_- + D^5_- + E^5_- + F^5_-)\Big\}(2\tau)\non\\
&-& \half\ \chi^{-,\text{(R)}}_{\text{st}}\ \Big\{ (\wh{A}^5_- + 
\wh{G}^5_- + \wh{H}^5_-
+ \wh{I}^5_- + \wh{J}^5_-)
\ +\ (\wh{B}^5_- + \wh{C}^5_- + \wh{D}^5_- + \wh{E}^5_- + \wh{F}^5_-)
\Big\}(2\tau)\ \Bigg]
\non\\
&=&\chi^{-,\text{(NS)}}_{\text{st}} \ (\NS^-_0\,+\,\NS^-_1) \ -\  
\chi^{-,\text{(R)}}_{\text{st}}\ (\RA_0^- + \RA_1^-)
\een

So adding eqns.\refb{q6} and \refb{q7}, we find the total KB amplitude in 
the orientifold of quintic\fn{In a supersymmetric theory the NSNS and RR part
of a KB amplitude cancel each other and hence the total KB amplitude vanishes
identically. We expect that both eqns. \refb{KBTOTAL} in section 
\ref{nss1.3} and \refb{q8} above vanish, though proving it requires certain
non-trivial identities involving Jacobi's $\vtheta$-functions.} 
\begin{eqnarray}\label{q8}
\text{KB}^{(k=3)^5}_{\text{total}} &=& (NS + \wt{NS}) - (R + \wt{R})\non\\
&=&  \Big(\chi^{+,\text{(NS)}}_{\text{st}}\ (\NS^+_0\,+\,\NS^+_1)\ +\  
\chi^{-,\text{(NS)}}_{\text{st}}\ (\NS^-_0\,+\,\NS^-_1)\Big)(2\tau)\non\\ 
&-& 
\Big(\chi^{+,\text{(R)}}_{\text{st}}\ (\RA_0^+ + \RA_1^+)\ +\ 
\chi^{-,\text{(R)}}_{\text{st}}\ (\RA_0^- + \RA_1^-)\Big)(2\tau)
\end{eqnarray}
As in case of $(k=1)^3$ model, the {\em different} crosscaps preserving 
{\em same} 
supersymmetry gives rise to {\em massive} characters.

From eqn.\refb{q8}, we find that the characters $(\NS^+_0 \ +\ \NS^+_1)$
appear with a plus sign, which implies implies that the
symmetric parts of the NSNS sectors are projected in. Note that
$\NS^+_0$ is the graviton
sector and this is correct. $\NS^+_1$ is the vector multiplet and this implies
that the scalars that appear from the NSNS sector are projected in. The
characters $(\RA^+_0 \ +\ \RA^+_1)$ in eqn.\refb{q8} appear with a minus 
sign. This implies that  the graviphoton and the vector in vector multiplet
are projected out. This is identical to what we found in section \ref{scas1}.

The next step would be check the conditions that are obtained from tadpole
cancellation. In this regard, we observe that the $(1,1,1,1,1)$ 
Recknagel-Schomerus states are the D-brane boundary states that should
be added to obtain vanishing of tadpoles. One can see that this is
indeed possible by studying the corresponding annulus amplitudes.

\sectiono{Conclusion}\label{con1}

In this paper we have studied the orientifolds of Calabi-Yau manifolds at
the Gepner point using techniques of rational conformal field theory. In 
particular, we have constructed the crosscaps for quintic. We
computed the Klein bottle amplitudes using these crosscaps for quintic.
The result we obtained from this abstract RCFT techniques has been 
verified against the geometric as well as SCFT results. To our 
satisfaction, they agree.

Obviously, this is only a first step. We did not discuss the
important unoriented open string sector which has to be included to cancel
RR tadpoles and extraction of the spectra. This can be extracted from the 
M\"obius strip amplitudes. This will be discussed in our next 
paper\cite{nextproone}. Other issues that we hope to address
include the case of even $k$ (not considered here for simplifying
the analysis), computation of intersection matrices via a Witten
index computation, extraction of RR charges. The case of even $k$ is 
interesting in that it has a richer structure of anti-holomorphic involutions.
This might lead to the appearance of chiral fermions in the spectra in addition
to non-abelian gauge symmetry. 

The case of type IIB orientifolds is interesting as well. In this case,
one deals with holomorphic involutions whose fixed points are holomorphic
submanifolds. The K\"ahler moduli which are complex become real after
orientifold projection. This implies that the large-volume $CY_3$, say the
quintic,  may
be separated  from the Gepner model by the conifold singularity (there will
be no way to ``go around" the singularity. At large volume, D-branes are
related to coherent sheaves, in general. Orientifolding at large volume,
involves extending the notion of duals of vector bundles to sheaves as well.
The use of $\pi$-stabilty to discuss the stability of D-branes\cite{stability}
in the
orientifold will require some modification. This has to take into account
that an unstable brane in the $CY_3$ case may become stable because its
decay product is projected out in the orientifold. This issues will be
discussed in a forthcoming paper\cite{SGTJ}.
The obvious extension of our construction
to the case of B-type crosscaps will also be discussed in the future.
In fact, one can use the Greene-Plesser construction\cite{GreeneP},
 where the mirror
$CY_3$ is obtained by further orbifolding of the Gepner model, to easily
write the B-type crosscap state by a trivial use of the orbifolding 
procedure of \cite{BH1} on the crosscaps that we have constructed.

\bigskip

\centerline {\bf Acknowledgments}

We acknowledge discussions with B. Acharya, I. Brunner, M. R. Douglas, 
B. Florea, C. Hofman, T. Jayaraman and H. Liu.
S.G. would like to thank the New High Energy Theory Center of Rutgers 
University for the invitation as well as 
hospitality during Summer 2002 during which this work was
started and the Theory Group at CERN where the manuscript was completed. 
We also would like to thank the organisers of PASCOS '2003
where these results were first reported\cite{Pascos}.

\addcontentsline{toc}{section}{Appendix}
\appendix{Notation for Boundary and Crosscap States in RCFT}\label{A1}

Let $J_n$, $\Jbar_n$ denote the modes of chiral and anti-chiral currents. 
then the definitions of boundary and crosscap states are :
\ben
\big(\,J_n\,+\,(-1)^{h_J}\,\Jbar_n\,\big)\ket{B} &=& 
0\,,\quad\hbox{Boundary}\\\non
\big(\,J_n\,+\,(-1)^{h_J\,+\,n}\,\Jbar_n\,\big)\ket{C} &=& 
0\,,\quad\hbox{Crosscap}\label{A.1}
\een
A basis for the solutions to eqn. \refb{A.1} are formed by {\em Ishibashi 
states}. They are given by
\ben\label{A.2}
\hbox{Boundary} : \ket{B_i}\rangle &=& 
\sum_I\,\ket{I}_i\,\ox\,U_B\,\ket{I}_{i^c}\\\non
\hbox{Crosscap} : \ket{C_i}\rangle &=& 
\sum_I\,\ket{I}_i\,\ox\,U_C\,\ket{I}_{i^c}\,,
\een
where $i$ labels a representation of the chiral algebra and $i^c$ its 
charge conjugate. The sum is over all states in a given representation. 
$U_B$ and $U_C$ are operators which satisfy :
\begin{equation}\label{A.3}
\Jbar_n\,U_B \,=\, (-1)^{h_J}\,U_B\,\Jbar_n\,;\qquad \Jbar_n\,U_C \,=\, 
(-1)^{h_J\,+\,n}\,U_C\,\Jbar_n
\end{equation}
Any boundary or crosscap state must be a linear combination of these 
Ishibashi states :
\begin{equation}\label{A.4}
\ket{B_a} \,=\, \sum_i\,B_{a\,i}\,\ket{B_i}\rangle\,;\qquad \ket{C} \,=\, 
\sum_i\,\Gamma_i\,\ket{C_i}\rangle
\end{equation}

\appendix{Constraints on Open string amplitudes}\label{A2}

\begin{itemize}
\item The coefficients $\MS^i_{\;a}$ satisfy extra condition\cite{PSS}, 
\viz
\begin{equation}\label{B.1}
\Big|\,\MS^i_{\;a}\,\Big|\,\le\,\AN^i_{\;a\,a}\,, 
\quad\MS^i_{\;a}\,=\,\AN^i_{\;a\,a} \mod 2
\end{equation}
This ensures that the open string sector has non-negative, integer state 
degeneracies in the sum $\dfrac{(\AN\,+\,\MS)}{2}$.
\item {\bf Completeness condition for the coefficients of Annulus 
amplitudes}

\smallskip
The annulus coefficients in direct channel are not all independent. They 
satisfy a set of polynomial equations:
\begin{eqnarray}\label{B.2} 
\AN_a^{\;\,i\,b}\,\AN_{b\,c}^{\;\,j}\,=\,\sum_k\,\NN^{i\,j}_{\;\,k}\,
\AN_{a\,c}^{\;\,k}
\\
\label{B.3}
\AN_{i\,a\,b}\,\AN^i_{\;\,c\,d}\,=\,\sum_i\,\AN_{i\,a\,c}\,\AN_{b\,d}^{\;\,i}
\end{eqnarray}
\end{itemize}

\medskip
\noi{\bf Comments :}
\begin{itemize}
\item Upper and lower boundary indices in eqns. \refb{B.2} and 
\refb{B.3} are to be distinguished in presence of oriented boundaries.
\item The matrix $(\AN_1)_{a\,b}\,=\,(\AN_1)^{a\,b}$ is a metric for 
boundary indices, as it follows from eqn. \refb{B.2} that
\ben\label{B.4}
\sum_b\,\AN_{i\,a\,b}\,\AN_1^{\;\,b\,c} 
&=&\AN_{i\,a}^{\;\,c}\,,\;\;\hbox{while}\non\\
(\AN_1)_{a}^{\;\,b} &=& \delta_a^{\;\,b}
\een
\item In diagonal models where $\AN$ coincides with $N$, eqs.\refb{B.2} 
and 
\refb{B.3} reduce to Verlinde algebra.
\item $\AN_{i\,a\,b}$ are in general {\em linearly dependent}, as
\be\label{B.5}
\sum_i\,\AN_{i\,\,a\,b}\,S_j^{\;\,i}\,=\,0\,,
\end{equation}
where the label $j$ in eq.\refb{B.5} is such a representation for which 
$Z_{j\,\jbar}\,=\,0$ in $\TO$.
\item {\bf Completeness conditions and Reflection Coefficients}
\smallskip

Define reflection coefficients as :
\begin{equation}\label{B.6}
\RE_{i\,a}\,=\,B_{i\,a}\,\sqrt{S_{i\,0}}
\end{equation}
Then the completeness conditions(eqns. \refb{B.2} and \refb{B.3}) are 
satisfied iff\cite{PSS,ZUBER},
\begin{equation}\label{B.7}
\sum_i\,\RE_{i\,a}\,\RE^{\;\ast}_{i\,b}\,=\,\delta_{a\,b}\,,\qquad 
\sum_a\,\RE_{i\,a}\,\RE^{\;\ast}_{j\,a}\,=\,\delta_{i\,j}
\end{equation}
\end{itemize}

\appendix{Formulae for simple currents}\label{A3}

Simple currents were discovered in refs.\cite{SY} in order to construct 
new modular invariants in rational conformal field theory. By definition, 
a {\em simple current} is a primary field of the algebra whose fusion with 
other primary fields produces only one primary field. This is the notion 
of being {\em  simple}. Thus if $J$ be such a simple current and 
$\Phi_{\,i}$ denote a generic primary field of the theory, by definition
\begin{equation}\label{C.1}
J\,\ox\,\Phi_{\,i}\,=\,\Phi_{\,k}
\end{equation}
Every conformal field theory has an obvious simple current, \viz the 
identity. There are examples of non-trivial simple currents in RCFTs. The 
very existence of a simple current in a RCFT, has some nice consequences;
first of all, $J$ will have its conjugate, $J^c$, such that
\begin{equation}\label{C.2}
J\,\ox\,J^c\,=\,\one
\end{equation}
Moreover, products of simple currents are simple currents. Thus powers of 
$J$ define an {\em orbit} of simple currents, all of which are different, 
unless one reaches the identity. Since the number of primary fields in a 
RCFT is finite, there must exist an integer $\N$, such that
\begin{equation}\label{C.3}
J^{\N}\,=\,\one
\end{equation}
If this is the smallest positive integer with such property, then $\N$ is 
called the {\em order} of the simple current.

The {\em monodromy charge} associated with a primary field, $\Phi_{\,i}$ 
with respect to 
the simple current $J$ is defined by computing the monodromy of 
$\Phi_{\,i}$ around the simple current $J$. If $Q_{\,J}(\Phi_{\,i})$ 
denote the monodromy charge of the field $\Phi_{\,i}$, then
\begin{equation}\label{C.4}
J(z)\,\Phi_{\,i}(w)\,\sim\,(z\,-\,w)^{\,-\,Q_{\,J}(\Phi_{\,i})}\,
\big(\,J\,\ox\,\Phi_{\,i}\,\big)(w)\,,
\end{equation}
so that
\begin{equation}\label{C.5}
Q_{\,J}(\Phi_{\,i})\,=\, 
h_{\,J}\,+\,h_{\,\Phi_{\,i}}\,-\,h_{\,J\,\ox\,\Phi_{\,i}} \mod 1
\end{equation}
$Q_{\,J}(\Phi_{\,i})$ satisfies the following property;
\begin{equation}\label{C.6}
Q_{\,J}(\Phi_{\,i}\,\Phi_{\,i}^{\prime})\,=\,Q_{\,J}(\Phi_{\,i})\,+\,
Q_{\,J}(\Phi_{\,i}^{\prime}) \mod 1
\end{equation}
The monodromy charge of the simple current is defined as
\begin{equation}\label{C.7}
Q_{\,J}(J)\,=\,\frac{\tilde{r}}{\N} \mod 1\,,
\end{equation}
where $\tilde{r}$ is defined modulo $\N$. Using 
$\Phi_{\,i}\,=\,J^{\,n\,-\,1}$ in eqn. \refb{C.5} and eqn. \refb{C.6}, one 
gets a recursion relation for the conformal weights of the currents,
\begin{equation}\label{C.8}
h_{\,J^n}\,=\,h_{\,J^{\,n\,-\,1}}\,+\,h_{\,J}\,-\,(n\,-\,1)\,Q_{\,J}(J) 
\mod 1
\end{equation}

\appendix{${\cal N}=2$ Minimal models}\label{A4}

The $k$-th $N=2$ minimal models  has central charge
$c=3k/(k+2)$. It can also be realised as the coset
$SU(2)_k\times U(1)_2/U(1)_{k+2}$. 
The primary fields of the model are specified by two integers
$(l,m)$. However, it is useful to split
Verma module of  a given ${\cal N}=2$ representation
into two sectors, those with even or odd worldsheet fermion
number. These sectors are distinguished by an extra label, $s$.
Thus, one has three integers $(l,m,s)$ ($l=0,1,\ldots,k$)
subject to the constraint that $l+m+s$ is even and
the representation $(l,m)=(l,m,s)\oplus (l,m,s+2)$.
Even $s$ refers to the NS sector and odd $s$ refers to the R sector
fields. 
The labels have the following field identification given by 
\begin{equation}\label{D.1a}
(l,\,m,\,s) \sim (l,\,m\,+\,2\,k\,+\,4,\,s)
 \sim (l,\,m,\,s\,+\,4)
\sim (k\,-\,j,\,m\,+\,k\,+\,2,\,s\,+\,2)
\end{equation}
The periodicity conditions in the $m$ and $s$ labels can be fixed by
choosing $(m,s)$ values as given below
\begin{equation}\label{D.1}
m=-(k+1),-k,\cdots,(k+2)\ \quad,\quad
 s=-1,0,1,2\ \quad, 
\end{equation}

The dimension $h$ and $U(1)$ charge $q$ of the fields are given by
\begin{eqnarray}\label{D.4}
h_{l,m,s}&=& \Delta_{l,m,s}\quad\mod 1
\nonumber \\
q_{l,m,s}&=&\frac{m}{k+2} - \frac{s}{2}\quad \mod 2
\end{eqnarray}
where $\Delta_{l,m,s}\equiv \frac{l(l+2) - m^2}{4(k+2)}~+~\frac{s^2}{8}$.

The $k$-th minimal model has a $\BZ_{k+2}\times \BZ_2$ discrete
symmetry. The action of the discrete symmetry on the fields is given by
\begin{eqnarray}\label{D.5}
g\cdot \Phi_{l,m,s} = e^{\frac{2\pi i m}{k+2}}\ \Phi_{l,m,s}\quad,
\\
h \cdot \Phi_{l,m,s} = (-)^s\ \Phi_{l,m,s}\quad,
\end{eqnarray}
where $g$ and $h$ generate the $\BZ_{k+2}$ and $\BZ_2$ respectively.

For our purposes, we need 
a set of labels in the minimal model that provide a single representative
after taking into account all identifications.
This is given by
the set ${\sf FR}={\sf FR}_{\rm NS}\cup {\sf FR}_{\rm R}$, where
\begin{equation}\label{D.5a}
\begin{split}
{\sf FR}_{NS} \equiv \left\{ (l,m,s)~|~0\leq l \leq k\ ;\ |m|\leq l\ ;\ 
s=0,2\ ; (l+m)= {\rm even} \right\} \\
{\sf FR}_{R} \equiv \left\{ (l,m,s)~|~0\leq l \leq k\ ;\ |m-1|\leq l\ ;\ s=\pm1\ ; (l+m)= {\rm odd}
\right\}
\end{split}
\end{equation}
We will need the exact conformal weights of these fields. Let 
$\widetilde{\sf FR}$ be the set ${\sf FR}$ without the elements
$\big((0,0,2),(l,l+1,-1)\big)$ ($l=0,\ldots,k$). The exact conformal weights
are given by
\begin{equation}\label{D.5b}
h_{l,m,s}=\begin{cases}\Delta_{l,m,s}&\text{for}\ 
(l,m,s)\in\widetilde{\sf FR}\\ 
                        \Delta_{l,m,s}+1 & \text{for}\ (l,m,s)\in
\big((0,0,2),(l,l+1,-1)\big)\ \big(l=0,\ldots,k\big) 
\end{cases}
\end{equation}

The crucial part in the computation of the $P$-matrix is to take the 
square-root of the $T$-matrix. We would like to write a formula that 
is compatible with the various identifications of the $(l,m,s)$. We write
it as
\begin{equation}\label{D.5c}
\big(\sqrt{T}\big)_{l,m,s; l,m,s}=\widehat{\sigma}_{l,m,s}\ e^{\pi i \Delta_{l,m,s}}\ ,
\end{equation}
where $\widehat{\sigma}_{l,m,s}$ is a sign.
Recall that $\Delta_{l,m,s}$ gives the weight of the primaries modulo 1. 
(We defined $h_{l,m,s}$ to be the exact weight.)
Consistency with the various identifications, implies
that the $\widehat{\sigma}_{l,m,s}$ must satisfy:
\begin{eqnarray}\label{sigmarels}
\widehat{\sigma}_{k-l,m-k-2,s-2}&=&(-)^{\frac{-l-m+s}2}\ \widehat{\sigma}_{l,m,s}\ ,\nonumber \\
\widehat{\sigma}_{l,m+2k+4,s}&=&(-)^{m+k}\ \widehat{\sigma}_{l,m,s}\ , \\
\widehat{\sigma}_{l,m,s+4}&=&(-)^{s}\ \widehat{\sigma}_{l,m,s}\ . \nonumber
\end{eqnarray}
In principle, any choice of $\widehat{\sigma}$ that satisfies eqn. 
(\ref{sigmarels}) should be acceptable. We however, will choose
it such that $\big(\sqrt{T}\big)_{l,m,s; l,m,s}=\ e^{\pi i h_{l,m,s}}$. This is 
achieved if we choose:
\begin{equation}\label{sigmadef}
\wh{\sigma}_{l,m,s}=
\begin{cases} 1&\text{for}\ (l,m,s)\in \widetilde{\sf FR} \\
                            -1& \text{for}\ (l,m,s)\in\big((0,0,2),(l,l+1,-1)\big)\ \big(l=0,\ldots,k\big) 
\end{cases}
\end{equation}
The values of $\widehat{\sigma}$ outside ${\sf FR}$ is then recursively
defined by the relations in eqn. (\ref{sigmarels}).

\subsection{$S$ and $P$ matrices for $U(1)_k$}\label{A4s1}

The primaries of the $U(1)_k$ are labelled by an integer
$m$ (defined mod $2k$) for which we take the standard range
${\sf SR}_k\equiv\{-k+1,-k+2,\ldots,k\}$ with weight $h_m=m^2/4k$.
The $S$ and $P$ matrices for the $U(1)_k$ are
\begin{eqnarray}\label{D.5d}
S_{m\,n} &=& \frac1{\sqrt{2k}} \exp\left[ -\frac{\pi i mn}{k}\right] \\
P_{m\,n} &=& \frac1{\sqrt{k}} \nu^{(k)}_m\nu^{(k)}_n\exp\left[ -\frac{\pi i mn}{2k}\right]
\ \delta^{(2)}_{m+n+k}
\end{eqnarray}
where we define $\nu^{(k)}_m$ as follows:
\begin{equation}\label{nudef}
\nu^{(k)}_m =\begin{cases}
                1 & {\rm for}\ m\in{\sf SR}_k\\
               (-)^{m+k} & {\rm for}\ (m\pm 2k)\in{\sf SR}_k
\end{cases}
\end{equation}
with the periodicity $\nu^{(k)}_{m+4k}=\nu^{(k)}_m$. The identity
$\nu^{(k)}_{m+2k} = (-)^{m+k}\nu^{(k)}_{m}$ holds for  all $m$.
Note that
the addition of the signs given by $\nu^{(k)}_m$ enables us to let
values of $m$ go outside the range ${\sf SR}_k$.

Using the expressions for the $S$ and $P$ matrices that we gave earlier,
one can show that
\begin{eqnarray}\label{D.5e}
{\NN^{(k)}_{m_1\, m_2}}^{m_3} &=& \delta^{(2k)}_{m_1 + m_2 - m_3} \\
{Y^{(k)}_{m_1\, m_2}}^{m_3} &=& \nu^{(k)}_{m_2}\nu^{(k)}_{m_3}\
\delta^{(2)}_{m_2+m_3} \left[
\delta^{(2k)}_{\frac{2m_1+m_2-m_3}2} + e^{\pi i(k+m_2)}\
\delta^{(2k)}_{\frac{2m_1+m_2-m_3+2k}2} \right]
\end{eqnarray}
Properties of the $Y$-tensor: ${Y_{(m_1+2k)\, m_2}}^{m_3}={Y_{m_1\,
m_2}}^{m_3}$

\subsection{$S$ and $P$ matrices for $SU(2)_k$}\label{A4s2}

The $S$ and $P$ matrices for the $SU(2)_k$ WZW model are
\begin{eqnarray}\label{D.5f}
{\cal S}_{L\,\tilde{L}} &=& \sqrt{\frac{2}{k+2}}\
\sin\big(L,\tilde{L}\big)_k \\
{\cal P}_{L\,\tilde{L}} &=& \frac{2}{\sqrt{k+2}}\
\sin\frac12\big(L,\tilde{L}\big)_k 
\ \delta^{(2)}_{L+\tilde{L}+k}
\end{eqnarray}
where $\big(l,l'\big)_k \equiv \left(\frac{\pi(l+1)(l'+1)}{k+2} \right)$.

The $\NN$-tensor for this case is
\begin{equation}\label{D.5g}
{{\cal N}_{L\,\tilde{L}}}^{l} =\begin{cases} 1 & |L-\tilde{L}|\leq l
\leq {\rm min}\{L+\tilde{L},2k-L-\tilde{L}\} \\
0 & {\rm otherwise} \end{cases} \ .
\end{equation}

\subsection{Y-Tensor elements for $SU(2)_k$}\label{A8}
\medskip

For convenience we list here the important components of the 
$SU(2)_k$ $Y$-tensor.
\begin{equation}\label{H.1}
\begin{array}{rcr}
Y^0_{l\ 0} = (-1)^l\,,\quad Y^k_{L, k - l} = \NN^l_{L\ L}\,,\\
Y^k_{l\ 0} = \NN^0_{l\ k - l} = \delta_{l,\  k -l}
\end{array}
\end{equation}

\subsection{$S$ and $P$ matrices for the minimal model}\label{A4s3}

Consider the following product of the $P$-matrices of $SU(2)_k$, $U(1)_{k+2}$
and $U(1)_{2}$ respectively.
\begin{equation}\label{pmatrix}
\widehat{P}_{LMS\ \tilde{L}\tilde{M}\tilde{S}} \equiv
{\cal P}_{L\, \tilde{L}} \times P^{*(k+2)}_{M\,\tilde{M}}\times P^{(2)}_{S\,\tilde{S}}
\end{equation}
The $S$-matrix and
the $P$-matrix of the $k$-th minimal model is then given by
\begin{eqnarray}\label{fullpmatrix}
\shoveright{S_{LMS\ \tilde{L}\tilde{M}\tilde{S}} = \frac1{\sqrt2(k+2)}\
\sin(l,l')_k \exp \left(\frac{i \pi mm'}{k+2}\right)
\exp\left(- \frac{i \pi ss'}2 \right)\hspace*{1.5in}}\\
\shoveright{
P_{LMS\ \tilde{L}\tilde{M}\tilde{S}} = \frac12 \widehat{\sigma}_{\tilde{L}\tilde{M}\tilde{S}}\left[
\widehat{\sigma}_{LMS}\
\widehat{P}_{LMS\ \tilde{L}\tilde{M}\tilde{S}}
+ \widehat{\sigma}_{(k-L)(M+k+2)(S+2)}\
\widehat{P}_{(k-L)(M+k+2)(S+2)\ \tilde{L}\tilde{M}\tilde{S}}  \right]
}
\end{eqnarray}
Using the conditions in eqn. 
(\ref{sigmarels}), one can show that the $P$-matrix is
unchanged under all identifications.

\subsection{$S$ and $P$ matrices for $SO(2d)_1$}\label{A4s4}

The four irreps of $SO(2d)$ are the
scalar, vector, spinor and conjugate spinor ($O,V,S,C$) representations.
We will represent them by the label $s_0=0,2,-1,1$.
The $S$ and $P$ matrices are (see the last ref. in\cite{PSS}) 
The $T$ matrix is 
\begin{equation}
\begin{split}
T&=e^{-\pi i d/12}\text{diag}\big(1,-1, e^{\pi i d/4},e^{\pi i d/4}\big)\\[3pt]
S&= \frac12\ \begin{pmatrix} 1 & 1 & 1 & 1 \\
                   1 & 1 &-1 &-1 \\
                   1 &-1 & i^{-d} & -i^{-d} \\
                   1 &-1 & -i^{-d} & i^{-d}
   \end{pmatrix}
\quad,\quad
P=\begin{pmatrix} c & s & 0 & 0 \\
            s & -c& 0 & 0 \\
            0 & 0 & \zeta c & i \zeta s \\
            0 & 0 & i \zeta s & \zeta c 
           \end{pmatrix}
\end{split}
\end{equation}
where $s=\sin(d\pi/4)$, $c=\cos(d\pi/4)$ and $\zeta=e^{-i d\pi/4}$.

\subsection{${\cal N}=2$ characters}\label{A4s5}

For a given representation $p$ of the ${\cal N}=2$ algebra, the character is
defined as
\begin{equation}\label{D.6}
\chi_p\left(\tau, z, u \right)~=~e^{-2 i \pi u}\ {\rm Tr}_p\ \Big[e^{2 i
\pi z 
J_0}\ e^{2 i \pi \tau (L_0-\frac{c}{24})}\Big]
\end{equation}
where the trace runs over the particular representation denoted by
$p$ and $u$ is an arbitrary phase. 
The characters of the $N=2$ minimal models are defined in terms of 
the level-$k$ theta functions $\Theta_{m,k}(\tau,z,u)$ defined in appendix 
\ref{A4s5} and characters of a related parafermionic
theory  $c^l_m(\tau)$ as: 
\begin{equation}\label{D.7}
\chi_{l,m}^{(s)}\left(\tau, z, u \right) = \sum_{t\,\in\,\Z_k}
c^{\,l}_{m+4t-s} (\tau)\ 
\Theta_{2m+(4t-s)(k+2),2k(k+2)}\left(\frac{\tau}{2}, 
\frac{z}{k\,+\,2}, u\right)\quad.
\end{equation}
The characters $\chi_{l,m}^{(s)}$ have the property that they are invariant
under $s \rightarrow s+4$ and $m \rightarrow m+2(k+2)$ and are zero 
if $l+m+s \neq 0$ mod $2$. For practical purpose, there is another useful 
formula(equivalent to eqn.\refb{D.7}) for the characters\cite{RY}
\begin{equation}\label{D.7prime}
\chi_{l,m}^{(s)}\left(\tau, z, u \right) = \sum_{m'\,=\,-k+1}^k\;
c^{\,l}_{m'}(\tau)\;\Theta_{\,m'(k+2) - mk + 2\,s\,k\,,\,k(k+2)}\;\Big(
\frac{\tau}{2}\,,\,\frac{z}{k+2}\,,\,u\Big)
\end{equation}
$c^{\,l}_m(\tau)$ are the famous string functions introduced by Kac and 
Peterson\cite{KP}. We list their symmetry properties below :
\begin{equation}\label{D.7pp}
\begin{array}{lcl}
c^l_m\,=\,c^l_{m + 2k\Z}\,=\,c^l_{-m}\,=\,c^{\,k - l}_{k - m}\\
c^l_m \,=\, 0\quad\text{if}\;\; l - m \ne 0 \mod 2
\end{array}
\end{equation}
The level-$k$ theta function is defined as :
\ben\label{E.1}
\Theta_{m\,,\,k}\big(\tau\,,\,z\,,\,u\big) &=& 
e^{-\,2\pi i u}\;\sum_{l\,\in\,Z}\;e^{2\pi i \tau 
k\,(l + \frac{m}{2k})^2\,+\,2\pi i\,z(l + \frac{m}{2k})}\non\\
&=& e^{-\,2\pi i u}\;\sum_{l\,\in\,Z}\;q^{k\,(l + \frac{m}{2k})^2}\;
y^{k\,(l + \frac{m}{2k})}\non\\
&=& e^{-\,2\pi i u}\ q^{\frac{m^2}{4k}}\ y^{\frac m2}\ 
\vtheta_3(kz + m\tau\bigl\lvert 2k\tau)\,,
\een
where $q\,=\,e^{2\pi i\tau}$ and $y\,=\,e^{2\pi i\, z}$ and we have 
used the following definition of Jacobi's $\vtheta$-functions 
\begin{equation}\label{E.5}
\vtheta\beb a\\b\eeb\,(z\bigl\lvert\tau)\,=\,
\sum_{n\,\in\,\Z}\;q^{\half(n + \frac{a}{2})^2}\;e^{2\pi i(z + 
\frac{b}{2})(n + \frac{a}{2})}\,.
\end{equation}
Following Jacobi/Erdelyi's notation, we have
\ben\label{E.7}
\vtheta_1 &=& \vtheta\beb 1\\1\eeb\,,\quad 
\vtheta_2 \,=\, \vtheta\beb 1\\0\eeb\,,\\
\vtheta_3 &=& \vtheta\beb 0\\0\eeb\,,\quad
\vtheta_4\,=\,\vtheta\beb 0\\1\eeb
\een
Level-$k$ theta function has the 
following symmetry property :
\begin{equation}\label{E.2}
\Theta_{m + 2k\,,\,k}\,=\,\Theta_{m\,,\,k}
\end{equation}
The Weyl-Kac character formula relates level-$k$ $SU(2)$ characters to level-
$k$ theta functions through the following identity : 
\begin{equation}\label{E.4}
\chi_l^k\,=\,\frac{\Theta_{l + 1\,,\,k + 2}\,-\,\Theta_{-l - 1\,,\,k + 2}}
{\Theta_{1\,,\,2}\,-\,\Theta_{-1\,,\,2}}
\end{equation}

By using the properties of the theta functions, 
the modular transformation of the minimal model characters is found to be
\begin{equation}\label{D.8}
\chi_{l,m}^{(s)} \left( -\frac1{\tau},0,0 \right) = C
\sum_{l',m',s'} \sin(l,l')_k \exp \left(\frac{i \pi mm'}{k+2}\right)
\exp\left(- \frac{i \pi ss'}2 \right) \chi_{l',m'}^{(s')}(\tau,0,0)
\end{equation}
where $(l,l')_k \equiv \left(\frac{\pi(l+1)(l'+1)}{k+2} \right)$
and $C=1/\sqrt2(k+2)$.

\appendix{Character Formulae of $k\,=\,1$ Minimal Model}\label{A6}
\medskip

\begin{itemize}
\item{\bf String Functions for $\mathbf{k\,=\,1}$} :

For $k\,=\,1$, due to its symmetry properties the only non-trivial string 
functions is $c^0_0(\tau)\,=\,c^1_1(\tau)$\cite{KP,GEPNER}. It is 
actually
\begin{equation}\label{F.3}
c^0_0(\tau)\,=\,c^1_1(\tau)\,=\,\frac{1}{\eta(\tau)}
\end{equation}

\item {\bf Formulae for the NS-sector Characters} :

The supersymmetric characters are defined as :
\ben\label{F.3a}
A_{\pm}(2\tau,z) &=& (\chi^0_{00} \pm \chi^2_{00})(2\tau,z)\,,\quad
B_{\pm}(2\tau,z) \,=\, (\chi^0_{11} \pm \chi^2_{11})(2\tau,z)\non\\
C_{\pm}(2\tau,z) &=& (\chi^0_{1-1} \pm \chi^2_{1-1})(2\tau,z)
\een
The characaters $A_-$, $B_-$ and $C_-$ appear in the $\wi$-twisted character
$\wt{NS}$. The formulae of $A_+$, $B_+$ and $C_+$ interms of theta functions 
are:
\ben\label{F.4}
A_+(2\tau,z) &=& 
\frac{1}{\eta(2\tau)}\;
\Theta_{0\,,\,3}\big(\tau\,,\,\frac{z}{3}\big)\,=\,
\frac{1}{\eta(2\tau)}\;\vtheta_3(z\bigl\lvert 6\tau)\,,\non\\
& & \\
B_+(2\tau,z) &=& 
\frac{1}{\eta(2\tau)}\;
\Theta_{2\,,\,3}\big(\tau\,,\,\frac{z}{3}\big)\,=\,
q^{\frac{1}{3}}\;y^{\frac{1}{3}}\;\frac{1}{\eta(2\tau)}\;
\vtheta_3(z + 2\tau\bigl\lvert 6\tau)\,,\non\\
& & \\
C_+(2\tau,z) &=& 
\frac{1}{\eta(2\tau)}\;
\Theta_{4\,,\,3}\big(\tau\,,\,\frac{z}{3}\big)\,=\,
\frac{1}{\eta(2\tau)}\;
\Theta_{-2\,,\,3}\big(\tau\,,\,\frac{z}{3}\big)\,=\,
q^{\frac{1}{3}}\;y^{-\frac{1}{3}}\;\frac{1}{\eta(\tau)}\;
\vtheta_3(z - 2\tau\bigl\lvert 6\tau)\,,\non\\
& & 
\een
\item{\bf Formulae for R-sector Characters} :

The supersymmetric characters in R-sectors are :
\ben\label{F.4a}
\wh{A}_{\pm}(2\tau,z) &=& (\chi^1_{01} \pm \chi^{-1}_{01})(2\tau,z)\,,\quad
\wh{B}_{\pm}(2\tau,z) \,=\,(\chi^1_{12} \pm \chi^{-1}_{12})(2\tau,z)\non\\
\wh{C}_{\pm}(2\tau,z) &=& (\chi^1_{10} \pm \chi^{-1}_{10})(2\tau,z)
\een
The characters $\wh{A}_-$, $\wh{B}_{-}$ and $\wh{C}_{-}$ appear in 
$\wi$-twisted character $\wt{R}$.

The R-sector characters are obtained by spectral flowing the NS-sector 
characters by an amount $\eta\,\to\,\eta +\ \half$ ($\imply\;\;z\,\to\,z +
\frac{\tau}{2}$) and using the property of 
$\Theta_{m\,,\,3}(\frac{\tau}{2}\,,\,\frac{z}{3})$ under spectral flow :
\begin{equation}\label{F.5}
\Theta_{m\,,\,3}\Big(\frac{\tau}{2}\,,\,\frac{z}{3}\Big)\;\xto{z\to z + 
\frac{\tau}{2}}\;q^{-\frac{1}{24}}\;y^{-\frac{1}{6}}\;
\Theta_{m + 1\,,\,3}\Big(\frac{\tau}{2}\,,\,\frac{z}{3}\Big)\;
\xto{z\to z + \frac{\tau}{2}}\;q^{-\frac{1}{6}}\;y^{-\frac{1}{3}}\;
\Theta_{m + 2\,,\,3}\Big(\frac{\tau}{2}\,,\,\frac{z}{3}\Big)
\end{equation}
The formulae for $\wh{A}_+$, $\wh{B}_+$ and $\wh{C}_+$ are given below :
\ben\label{F.6}
\wh{A}_+(2\tau,z) &=& 
\frac{1}{\eta(2\tau)}\;\Theta_{\,1\,,\,3}
\big(\tau\,,\,\frac{z}{3}\big)\,=\,\frac{1}{\eta(2\tau)}\;
q^{\frac{1}{12}}\;y^{\frac{1}{6}}\;
\vtheta_3\big(z + \tau\bigl\lvert 6\tau\big)\,,\non\\
& & \\
\wh{B}_+(2\tau,z) &=& 
\frac{1}{\eta(2\tau)}\;\Theta_{\,3\,,\,3}
\big(\tau\,,\,\frac{z}{3}\big)\,=\,\frac{1}{\eta(2\tau)}\;
q^{\frac{3}{4}}\;y^{\frac{1}{2}}\;
\vtheta_3\big(z + 3\tau\bigl\lvert 6\tau\big)\,,\non\\
& & \\
\wh{C}_+(2\tau,z) &=& 
\frac{1}{\eta(2\tau)}\;\Theta_{\,5\,,\,3}
\big(\tau\,,\,\frac{z}{3}\big)\,=\,
\frac{1}{\eta(2\tau)}\;\Theta_{\,-1\,,\,3}
\big(\tau\,,\,\frac{z}{3}\big)\,=\,
\frac{1}{\eta(2\tau)}\;q^{\frac{1}{12}}\;y^{-\frac{1}{6}}\;
\vtheta_3\big(z - \tau\bigl\lvert 6\tau\big)\,,\non\\
& & 
\een
\end{itemize}

\medskip

\appendix{Character Formulae and Spectral Flow Invariant Orbits 
of $k\,=\,3$ Minimal Model}\label{A7}
\medskip

\begin{itemize}
\item {\bf String Functions} :

There are four independent string functions at the level $k\,=\,3$. They 
are \newline
$\{c^0_0(\tau)\,,\,c^1_1(\tau)\,,\,c^2_0(\tau)\,,\,c^3_1(\tau)\}$. 
Their expressions had been given in Kac-Peterson's paper\cite{KP} 
\ben\label{G.4}
c^1_1(\tau) &=& \frac{q^{\frac{3}{40}}\;\prod_{n\ne \pm 1 \mod 5}\;
(1 - q^{3n})}{\eta(\tau)^2}\,,\\
c^3_1(\tau) &=& \frac{q^{\frac{27}{40}}\;\prod_{n\ne \pm 2 \mod 5}\;
(1 - q^{3n})}{\eta(\tau)^2}\,,\\
(c^0_0\,-\,c^3_1)(\tau) &=& \frac{q^{\frac{1}{120}}\;\prod_{n\ne \pm 1 
\mod 5}\;(1 - q^{\frac{n}{3}})}{\eta(\tau)^2}\,,\\
(c^1_1\,-\,c^2_0)(\tau) &=& \frac{q^{\frac{3}{40}}\;\prod_{n\ne \pm 2
\mod 5}\;(1 - q^{\frac{n}{3}})}{\eta(\tau)^2}
\een

\item {\bf Spectral flow} :

We list the spectral flow of all NS-sector characters in the table 5
below.
\renewcommand{\arraystretch}{1.2}
\begin{table}\label{G.1}
\bec
\begin{tabular}[!h]{||c|c|c|c|c|c||}\hline
Labels for & $l$ & $m$ & $h$ & $q$ & Spectral flow\\
the characters & & & & & $(\eta\,\to\,\eta\,+\,1)$\\\hline
$A_{\pm}$ & 0 & 0 & 0 & 0 & $A_{\pm}\,\to\,J_{\pm}$\\\hline
$B_{\pm}$ & 1 & $-$1 & $1/10$ & $- 1/5$ & $B_{\pm}\,\to\,F_{\pm}$\\
$C_{\pm}$ &  & 1 & $1/10$ & $1/5$ & $C_{\pm}\,\to\,B_{\pm}$\\\hline
$D_{\pm}$ &  & $-$2 & $1/5$ & $- 2/5$ & $D_{\pm}\,\to\,C_{\pm}$\\
$E_{\pm}$ & 2 & 0 & $2/5$ & 0 & $E_{\pm}\,\to\,D_{\pm}$\\
$F_{\pm}$ & & 2 & $1/5$ & $2/5$ & $F_{\pm}\,\to\,E_{\pm}$\\\hline
$G_{\pm}$ & & $-$3 & $3/10$ & $- 3/5$ & $G_{\pm}\,\to\,A_{\pm}$\\
$H_{\pm}$ & 3 & $-$1 & $7/10$ & $- 1/5$ & $H_{\pm}\,\to\,G_{\pm}$\\
$I_{\pm}$ & & 1 & $7/10$ & $1/5$ & $I_{\pm}\,\to\,H_{\pm}$\\
$J_{\pm}$ & & 3 & $3/10$ & $3/5$ & $J_{pm}\,\to\,I_{\pm}$\\\hline
\end{tabular}
\eec
\caption{The representations and their spectral flow for $k\,=\,3$ minimal
model.}
\end{table}

\item{\bf Formulae for the characters} :

The NS-sector characters of $k=3$ minimal model and their spectral flow are
given in the table 5. The formulae of ten characters 
in NS-sectors 
\ben\label{G.1a}
A_{\pm}(2\tau,z) &=& (\chi^0_{00} \pm \chi^2_{00})(2\tau,z)\,,\quad
B_{\pm}(2\tau,z) = (\chi^0_{1-1} \pm \chi^2_{1-1})(2\tau,z)\non\\
C_{\pm}(2\tau,z) &=& (\chi^0_{11} \pm \chi^2_{11})(2\tau,z)\,,\quad
D_{\pm}(2\tau,z) = (\chi^0_{2-2} \pm \chi^2_{2-2})(2\tau,z)\non\\
E_{\pm}(2\tau,z) &=& (\chi^0_{20} \pm \chi^2_{20})(2\tau,z)\,,\quad
F_{\pm}(2\tau,z) = (\chi^0_{22} \pm \chi^2_{22})(2\tau,z)\non\\
G_{\pm}(2\tau,z) &=& (\chi^0_{3-3} \pm \chi^2_{3-3})(2\tau,z)\,,\quad
H_{\pm}(2\tau,z) = (\chi^0_{3-1} \pm \chi^2_{3-1})(2\tau,z)\non\\
I_{\pm(2\tau,z)} &=& (\chi^0_{31} \pm \chi^2_{31})(2\tau,z)\,,\quad
J_{\pm}(2\tau,z) = (\chi^0_{33} \pm \chi^2_{33})(2\tau,z)
\een
The characters $A_+$, $B_+$, $C_+$, $D_+$, $E_+$, $F_+$, $G_+$, $H_+$, $I_+$ 
and $J_+$ can be obtained from the following level-15 theta function, by
changing the values of $m$ :
\begin{equation}\label{G.1b}
\Theta_{m\,,15}\big(\tau,\frac{z}{5},0\big) = q^{m^2/60}\ y^{m/10}\ 
\vtheta_3\big(3z + m\tau\bigl\lvert 30\tau\big)
\end{equation} 
\end{itemize}

The R-sector characters are
\ben\label{G.1c}
\wh{A}_{\pm}(2\tau,z) &=& (\chi^1_{01} \pm \chi^{-1}_{01})(2\tau,z)\,,\quad
\wh{B}_{\pm}(2\tau,z) = (\chi^1_{10} \pm \chi^{-1}_{10})(2\tau,z)\non\\
\wh{C}_{\pm}(2\tau,z) &=& (\chi^1_{12} \pm \chi^{-1}_{12})(2\tau,z)\,,\quad
\wh{D}_{\pm}(2\tau,z) = (\chi^1_{2-1} \pm \chi^{-1}_{2-1})(2\tau,z)\non\\
\wh{E}_{\pm}(2\tau,z) &=& (\chi^1_{21} \pm \chi^{-1}_{21})(2\tau,z)\,,\quad
\wh{F}_{\pm}(2\tau,z) = (\chi^1_{23} \pm \chi^{-1}_{23})(2\tau,z)\non\\
\wh{G}_{\pm}(2\tau,z) &=& (\chi^1_{3-2} \pm \chi^{-1}_{3-2})(2\tau,z)\,,\quad
\wh{H}_{\pm}(2\tau,z) = (\chi^1_{30} \pm \chi^{-1}_{30})(2\tau,z)\non\\
\wh{I}_{\pm(2\tau,z)} &=& (\chi^1_{32} \pm \chi^{-1}_{32})(2\tau,z)\,,\quad
\wh{J}_{\pm}(2\tau,z) = (\chi^1_{34} \pm \chi^{-1}_{34})(2\tau,z)
\een

\end{document}